\PassOptionsToPackage{noend}{algorithmic}
\documentclass{article} 
\usepackage{iclr2027_conference,times}
\makeatletter
\iclrfinalcopy
\makeatother

\usepackage{amsmath,amsfonts,bm}

\def\eqref#1{equation~\ref{#1}}

\def\1{\bm{1}}

\DeclareMathAlphabet{\mathsfit}{\encodingdefault}{\sfdefault}{m}{sl}
\SetMathAlphabet{\mathsfit}{bold}{\encodingdefault}{\sfdefault}{bx}{n}

\usepackage{hyperref}
\usepackage{url}

\usepackage{adjustbox}
\usepackage{tabularx}
\usepackage{inconsolata}

\usepackage{balance}

\usepackage{amsmath}
\usepackage{amssymb}
\usepackage{graphicx}
\usepackage[normalem]{ulem}
\usepackage{ulem}
\usepackage{subfigure}
\usepackage{booktabs}
\usepackage{amsmath}

\usepackage{pifont} 
\usepackage{algorithm}
\usepackage{algorithmic}
\usepackage[noend]{algorithmic}

\usepackage[most]{tcolorbox} 
\usepackage{makecell}
\usepackage{multirow} 
\usepackage{soul}
\usepackage{xcolor}
\definecolor{DeepGreen}{RGB}{0,128,0}

\sethlcolor{green!30}
\newcommand{\hlred}[1]{{\sethlcolor{red!25}\hl{#1}}}

\newcommand{\hlb}[1]{{\sethlcolor{cyan!15}\hl{#1}}}

\definecolor{swiftblue}{HTML}{2F5597}
\definecolor{swiftgreen}{HTML}{2E7D32}
\definecolor{swiftorange}{HTML}{F28C28}
\definecolor{swiftpurple}{HTML}{6A4C93}
\definecolor{swiftred}{HTML}{B23A48}
\definecolor{swiftgray}{HTML}{F7F7F7}

\newtcolorbox{swiftbox}[2]{
    enhanced,
    colback=#2!5,
    colframe=#2!85!black,
    colbacktitle=#2!85!black,
    coltitle=white,
    title={#1},
    fonttitle=\bfseries\footnotesize,
    boxrule=0.55pt,
    arc=1.5mm,
    left=1.2mm,
    right=1.2mm,
    top=1mm,
    bottom=1mm,
    before upper={\vspace{0.5mm}}
}

\title{Adaptive Co-Serving LLM Watermarking \\on Modern Inference Engines}

\author{
Kieu Dang\textsuperscript{1},
Phung Lai\textsuperscript{1}\thanks{Corresponding author: \texttt{lai@albany.edu}},
Ching-Yun Ko\textsuperscript{2},
Pin-Yu Chen\textsuperscript{2}
\\[0.5em]
\textsuperscript{1}State University of New York at Albany
\qquad
\textsuperscript{2}IBM Research 
}

\begin{document}

\maketitle

\begin{abstract}

 
Large language model (LLM) watermarking is important for ownership verification and intellectual property protection. However, existing approaches focus on algorithmic design while treating LLM inference engines as separate components. This separation often introduces auxiliary models or external tools, increasing latency and memory overhead while limiting the use of modern inference optimizations. As a result, a deployment gap remains: practical watermarking must preserve utility, detectability, and robustness while minimizing serving overhead. 

To bridge this gap, we propose \textsc{SWIFT}, a framework that co-designs LLM watermarking with modern inference infrastructure. \textsc{SWIFT} (i) integrates text generation and watermark construction within a shared LLM backend to reduce re-computation, improve cache reuse, and simplify system complexity; (ii) uses instruction-guided candidate generation with semantic protection to preserve factual consistency and meaning; (iii) executes watermarking concurrently with text generation through asynchronous co-serving and adaptive scheduling to reduce latency and improve resource utilization; and (iv) leverages vLLM optimizations to enhance serving efficiency.
Extensive experiments across domains and tasks show that \textsc{SWIFT} achieves strong utility, detectability, robustness, and downstream performance while substantially improving serving efficiency. Specifically, it achieves the highest utility score (4.87), 99.65\% detection accuracy compared with 76.65\% for the low-latency baseline SynthID, and up to 97.7\% detection under watermark removal attacks, with $5.9\times$ lower latency than the highest-utility baseline, SafeSeal. These results demonstrate the benefits of co-designing watermarking with   inference infrastructure for practical LLM serving. 



\end{abstract}

\section{Introduction}

As large language models (LLMs) \citep{OpenAI,claude,gemini} become widely deployed across domains, their outputs can be copied, transformed, or redistributed without  evidence of origin. This provenance gap makes it difficult to verify whether text was generated by a  model or service, especially after it leaves the provider. LLM watermarking \citep{kirchenbauer2023watermark,dathathri2024scalable,zhu2024duwak,munyer2024deeptextmark,yang2023watermarking} provides a practical solution by embedding imperceptible but detectable watermark signals in generated text, supporting origin verification and ownership attribution. For watermarking to be practical in   LLM services, however, it must  integrate efficiently into the inference engines that generate these outputs at scale.

Existing watermarking methods largely focus on watermark construction and detection algorithms \citep{kirchenbauer2023watermark,kuditipudi2023robust,dathathri2024scalable,yoo2024advancing}, while LLM inference engines have evolved independently to optimize latency, memory, and throughput \citep{li2023alpaserve,zhong2024distserve,Hu2025Deepserve}. This separation creates two  deployment gaps. First, watermarking is often implemented as an auxiliary pipeline, relying on  separate modules  or external natural language processing (NLP) tools 
such as spaCy~\citep{spaCy2020}, Stanza~\citep{stanza2020}, and  NLTK~\citep{bird2009nltk}   for named entity recognition (NER) and part-of-speech (POS) tagging,  
rather than sharing an engine with text generation  \citep{optimizing2024wouters24a,wang2025morphmark,hou2024token,huang2024waterpool}. Such fragmentation increases latency, memory usage, and system complexity. Second, serving optimizations such as asynchronous generation,    key-value cache (KV-cache) reuse, prefix sharing,  and  scheduling remain largely unexplored for watermarking.



 \begin{figure}[t]
      \centering    
      \includegraphics[scale=0.068]{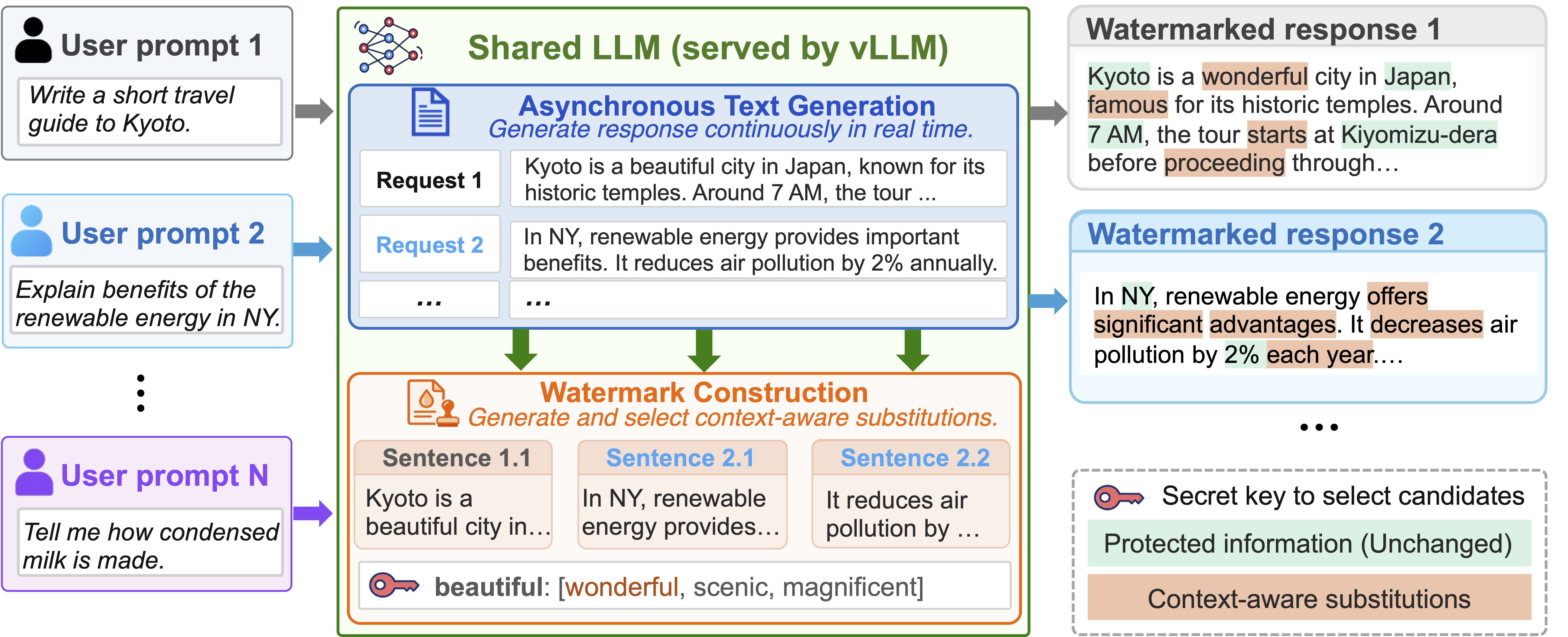} 
      \caption{\textbf{\textsc{SWIFT} example.} Multiple prompts are processed concurrently through a shared vLLM backend. For a Kyoto guide, \textsc{SWIFT} generates context-aware substitutions (e.g., ``beautiful'' $\rightarrow$ ``wonderful'') while preserving key information such as locations and times. Watermark construction runs asynchronously alongside text generation, enabling streaming with minimal latency overhead.}
      \vspace{-7.5pt}
      \label{fig:SWIFT_introduction}
 \end{figure}

These gaps motivate a  question: \textit{Can we design watermarking  as part of LLM inference engines, rather than as an auxiliary pipeline?} To address this question, we introduce \textsc{SWIFT} (Fig.~\ref{fig:SWIFT_introduction}), to our knowledge, the first high-utility \textbf{\underline{S}}erving-aware \textbf{\underline{W}}atermarking for \textbf{\underline{I}}nference-\textbf{\underline{F}}riendly \textbf{\underline{T}}ext generation framework that co-designs watermarking with modern inference engines. 
Built on vLLM \citep{kwon2023efficient}, a high-throughput LLM serving engine with efficient KV-cache management and request scheduling, \textsc{SWIFT} addresses the gaps through a unified co-serving architecture. First, it \textit{\textbf{co-serves text generation and watermark construction}} on the same LLM, inference engine, and GPU, eliminating separate   modules or NLP tools while reducing duplicated computation, memory overhead, and system complexity.
 For watermark construction, the shared LLM performs instruction-guided candidate generation, identifying watermarkable words,  protecting named entities and sensitive words, and generating context-aware substitutions under semantic and factual constraints. Second, \textsc{SWIFT} introduces an \textit{\textbf{adaptive co-serving scheduler}} that dynamically admits watermark requests based on queue length and waiting time pressure, preventing   backlogs while preserving resources for   text generation. This unified design further enables both workloads to share serving optimizations, including asynchronous execution, prefix caching, continuous batching, structured decoding, KV-cache management, and chunked prefill, allowing watermark computation to overlap with ongoing generation while reducing redundant computation and resource overhead.

Experiments across tasks and LLMs demonstrate that \textsc{SWIFT} improves watermark quality and serving efficiency over baselines, including KGW \citep{kirchenbauer2023watermark}, SynthID \citep{dathathri2024scalable}, and SafeSeal \citep{dang2026safeseal}. It achieves a utility score of 4.87 and 99.65\% detection accuracy  with 0.905 seconds (s) latency,  $5.9\times$ lower latency than the highest-utility baseline. \textsc{SWIFT} retains up to 97.7\% detection accuracy under removal attacks and generalizes to downstream tasks, including summarization, short reasoning, and technical medical text, without performance degradation. Qualitative analysis shows its substitutions better preserve factual consistency. Its key-conditioned detection   reaches  99.7\% with the correct key and near-zero detection with incorrect keys. From a serving perspective, \textsc{SWIFT} achieves high throughput of 528.7 tokens/s, while reducing KV-cache usage and attaining a 96.8\% prefix-cache hit rate. 
Overall, our approach combines high utility, robust detectability, and efficient serving, making LLM watermarking more practical.

\section{Background and Related Work}

\textbf{LLM Watermarking.} LLMs generate text autoregressively by decoding one token at a time. LLM watermarking  modifies this process or generated text to embed imperceptible but detectable watermark signals for provenance verification, with a generator embedding the watermark and a detector verifying its presence. Existing  methods broadly fall into two categories. \textbf{\textit{In-processing watermarks}} modify token selection during decoding \citep{zhao2024provable, liu2024adaptive, hu2024unbiased}. Integrated into text generation, these methods naturally support streaming but may not fully exploit modern serving optimizations  \citep{fernandez2025good}. Their per-token modifications may accumulate utility degradation over long outputs.  \textbf{\textit{Post-processing watermarks}} modify generated text,  often using auxiliary NLP tools or language models (LMs) \citep{topkkara2006words,chang2024postmark}. They can better preserve utility by   selecting   watermark locations and substitutions under semantic and contextual constraints  \citep{yang2022tracing,dang2026safeseal}, but  add  latency and memory overhead and are less suited to streaming.
These trade-offs suggest that watermarking performance depends on both the embedding algorithm and its integration with the underlying inference system.



\textbf{LLM Inference Engines and Serving Systems.} Modern LLM inference engines improve serving efficiency through KV-cache management, continuous batching, and request scheduling \citep{miao2025towards,sun2024llumnix}. As KV-cache memory grows with the number and length of active requests, it can limit serving concurrency \citep{jiang2026towards,mitzenmacher2025queueing}.
vLLM \citep{vllm_docs} addresses this bottleneck with PagedAttention \citep{kwon2023efficient}, which manages KV states in fixed-size blocks to reduce memory fragmentation and improve utilization. It also   supports continuous batching, prefix caching, 
asynchronous generation, and structured decoding. Other systems optimize complementary serving components: Orca introduces iteration-level scheduling \citep{yu2022orca}, SGLang improves cross-request KV reuse through RadixAttention \citep{zheng2024sglang}, and Sarathi-Serve uses chunked prefill to mitigate prefill-decode interference \citep{agrawal2024taming}. 
DistServe and Splitwise disaggregate prefill and decode execution \citep{zhong2024distserve,patel2024splitwise}, while others explore preemptive and dynamic request scheduling \citep{wu2023fast}.

 Despite these advances, watermarking has largely evolved independently of inference engine design. Most methods use frameworks such as Hugging Face Transformers \citep{kirchenbauer2023watermark,munyer2024deeptextmark,dang2026safeseal}, with limited  cache reuse, asynchronous co-execution, and adaptive scheduling. Existing vLLM integrations  mainly target in-processing watermarking \citep{vllm_watermark,vllm_watermarking_team}, while  post-processing methods with serving optimizations  remain underexplored. This gap limits watermarking from fully leveraging   LLM inference mechanisms.

\textbf{Challenges of Practical Watermarking.} Bridging this gap requires addressing two interconnected challenges. First, watermark construction is often implemented as a separate processing pipeline. In-processing methods introduce operations along the decoding path, whereas post-processing methods may rely on external NLP tools or auxiliary  LMs. These designs increase computation and deployment complexity. 
Second, watermarking introduces an additional workload that must coexist with latency-sensitive text generation. Static or uncoordinated execution can create watermark backlogs, compete for resources, and delay streaming outputs. 
Thus, practical watermarking is both an algorithmic and systems problem: it must preserve utility, detectability, and robustness while sharing inference resources efficiently. These challenges motivate \textsc{SWIFT}, which integrates watermark construction into a shared LLM backend and adaptively co-schedules it with ongoing text generation.

 \textbf{Threat Model.} We consider a proprietary LLM service provider where users can submit prompts and receive outputs via an API. A black-box adversary  can query the API and observe outputs but cannot access model  parameters, architecture, or internal states. The adversary seeks to weaken watermark detection while preserving the original content through watermark removal attacks such as paraphrasing, insertion,  synonym substitution, and rewriting \citep{zhang2023watermarks,lin2021towards}.

\section{SWIFT: Adaptive Co-Serving LLM Watermarking } \label{sec:swift}



\subsection{Design Objectives and System Overview}
\label{sec:system_overview}

The design of \textsc{SWIFT} follows four objectives: \textit{(i) Unified design}, minimizing external dependencies through a shared LLM backend; \textit{(ii) Utility preservation}, maintaining semantic and factual consistency through instruction-guided watermarking; \textit{(iii) Adaptive co-serving}, coordinating concurrent workloads to control watermark backlogs and maintain streaming responsiveness; and \textit{(iv) Serving efficiency}, leveraging inference optimizations to reduce computation, memory, and latency. 

Fig.~\ref{fig:SWIFT} and Alg.~\ref{alg:swift} (Appx.~\ref{app:Tech}) illustrate the   architecture of \textsc{SWIFT}, which integrates text   and watermark generation streams within a single vLLM  engine.  
As both streams share a backend, \textsc{SWIFT} loads the LLM once, enabling both workloads to leverage serving optimizations and reducing external dependencies. The   workloads execute asynchronously,  overlapping watermark processing  with ongoing text generation, while an adaptive co-serving scheduler coordinates their execution under varying demand. Together, these components balance watermark quality, detectability, and serving efficiency. The following subsections describe co-serving, scheduling, and watermark construction.

\subsection{Asynchronous Co-Serving of Text  and Watermark Generation}
\label{sec:dual_stream}

\textsc{SWIFT} co-serves text generation and watermark construction as two asynchronous inference streams sharing the same vLLM engine, LLM, and GPU. Given a prompt $x$, the text generation stream autoregressively decodes a response $y=(y_1,\ldots,y_T)$ $(T>0)$, producing tokens incrementally rather than waiting for the complete response. This enables streaming generation while allowing watermark processing to begin before text generation finishes, reducing the latency of conventional generate-then-watermark pipelines \citep{he2022protecting,hao2025post,yang2022tracing}. By reusing the same inference backend for both workloads, \textsc{SWIFT} eliminates separate watermarking engines, external NLP tools, and auxiliary LMs, reducing complexity and memory overhead.

 \begin{figure*}[t]
      \centering    
      \includegraphics[scale=0.056]{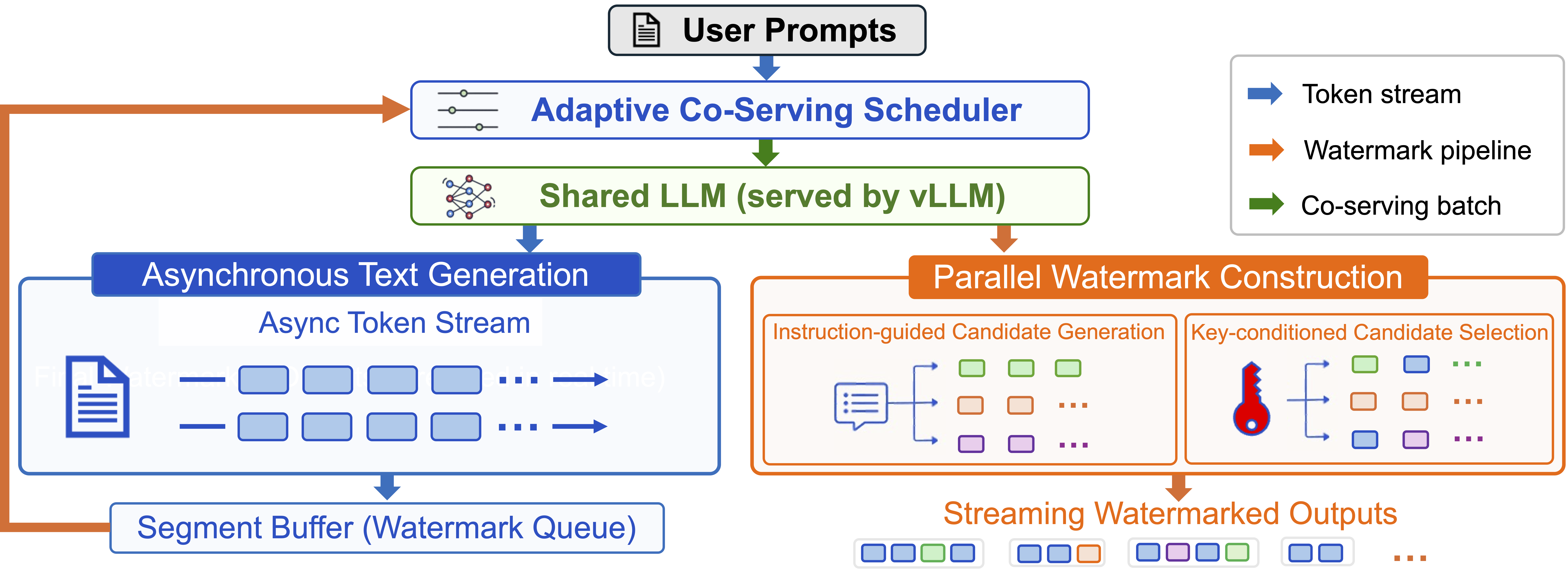} 
      \caption{Overview of \textsc{SWIFT}.}
      \vspace{-5pt}
      \label{fig:SWIFT}
 \end{figure*}

 To enable asynchronous execution, \textsc{SWIFT} maintains a segment buffer that collects generated tokens and identifies completed segments (typically sentences). At decoding step $t$, token $y_t$ is appended to the buffer. Upon detecting a sentence boundary, \textsc{SWIFT} extracts the completed segment $s_i$ and submits it as a   watermarking request to the shared LLM. Meanwhile, text generation continues decoding subsequent tokens $y_{t+1},y_{t+2},\ldots$ without waiting for watermarking to finish (Fig.~\ref{fig:SWIFT}). The adaptive co-serving scheduler coordinates watermark request admission alongside ongoing generation to balance both workloads, as detailed in Section~\ref{sec:adaptive_scheduler}. Once processed, watermarked segments are streamed to users in their original order, maintaining incremental output delivery.

Because both streams share the same LLM backend, \textsc{SWIFT} loads the model once and leverages vLLM serving optimizations to improve efficiency \citep{vllm_docs}. Prefix caching reuses KV states for shared watermark instructions, while continuous batching incorporates newly admitted watermark requests. Request prioritization manages latency-sensitive workloads, and chunked prefill mitigates prefill-decode interference. Shared KV-cache management and consistent model precision further reduce memory overhead and avoid redundant model representations or conversion costs.


\subsection{Adaptive Co-Serving Scheduler}
\label{sec:adaptive_scheduler}

\begin{figure*}[t]
      \centering    
      \includegraphics[scale=0.050]{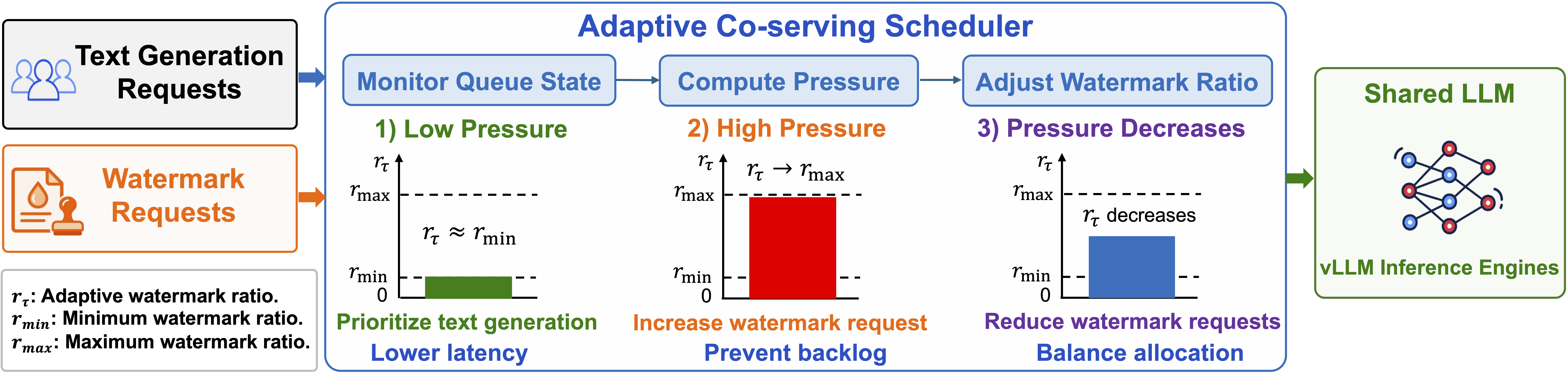} 
      \caption{Adaptive co-serving scheduler behaviors.} 
      \vspace{-5pt}
      \label{fig:adaptive_scheduler_overview}
 \end{figure*}

To coordinate text generation and watermark construction, \textsc{SWIFT} introduces an \textit{\textbf{adaptive co-serving scheduler}} that adjusts watermark request admission based on runtime workload pressure (Fig.~\ref{fig:adaptive_scheduler_overview}). The inference engine co-serves text generation and watermark requests, each occupying a sequence in the mixed serving batch (\textit{mixed-batch}). Rather than reserving a fixed portion of the batch for watermarking, the scheduler adapts watermark admission based on queue length and waiting time, balancing workloads while maintaining streaming responsiveness and resource efficiency.

\noindent \textbf{Scheduler Update and Adjustment.}
To  balance the two request types, the adaptive scheduler updates the watermark admission ratio $r_{\tau}$ every $K$ scheduling steps,
  where $\tau$ is a scheduling step:
\begin{equation}
r_{\tau}
=
r_{\min}
+
\left(r_{\max}-r_{\min}\right)p_{\tau},
\label{eq:adaptive_ratio}
\end{equation}
where $r_{\min}$ and $r_{\max}$ define the lower and upper bounds of the watermark admission ratio, and $p_{\tau}\in[0,1]$ represents normalized watermark   pressure. This formulation   adjusts watermark admission according to demand. 
Under low pressure ($p_{\tau}\approx 0$), $r_{\tau}$ remains close to $r_{\min}$, preserving more   capacity for text generation. As pressure increases ($p_{\tau}\rightarrow 1$),   $r_{\tau}$ rises toward $r_{\max}$ to admit more watermark requests and limit backlog accumulation.  Once pressure decreases, subsequent updates reduce $r_{\tau}$, returning capacity to text generation and maintaining streaming responsiveness.

\noindent \textbf{Watermark Pressure Estimation.} To obtain the normalized watermark pressure $p_{\tau}$, \textsc{SWIFT} monitors the first-in, first-out (FIFO) queue of completed segments awaiting watermarking. Let $q_{\tau}$ denote the watermark queue length (i.e., the number of pending watermark requests) and $w_{\tau}$ denote the waiting time of the oldest request.  
\textsc{SWIFT} combines $q_{\tau}$ and $w_{\tau}$ to estimate watermark pressure: 
\begin{equation}
p_{\tau}
=
\operatorname{clip}\left(
\max\left(
\frac{q_{\tau}}{B_{\max}},
\frac{w_{\tau}}{W_{\max}}
\right),
0,1
\right).
\label{eq:wm_pressure}
\end{equation}
Here, $B_{\max}$ and $W_{\max}$ denote the queue length and waiting time thresholds, respectively. The term $\frac{q_{\tau}}{B_{\max}}$ captures \emph{backlog pressure}, measuring pending requests relative to the queue threshold, while $\frac{w_{\tau}}{W_{\max}}$ captures \emph{waiting time pressure}, measuring the oldest request's delay relative to its threshold. These terms capture complementary conditions: a large queue indicates accumulating demand, while a long waiting time reveals delays even with few pending requests. Taking the maximum prioritizes the more urgent condition, while clipping $p_{\tau}$ to $[0,1]$ keeps $r_{\tau}$ within its prescribed limits.


\noindent \textbf{Adaptive Mixed-Batch Admission.} 
The scheduler translates $r_{\tau}$ into the target watermark capacity within the shared batch. Let $B$ denote the maximum number of concurrent sequences supported by the shared vLLM. The target watermark capacity is $B^{\mathrm{wm}}_{\tau}=\left\lfloor r_{\tau} B \right\rfloor$, with the remaining   capacity available for text generation. Thus, $r_{\tau}$ controls watermark request admission without modifying the decoding process. Both request types execute within the shared vLLM engine, leveraging serving optimizations while adapting concurrency to limit watermark backlogs and processing delays.


\subsection{Watermark Construction and Detection}
\label{sec:watermark_generation}

To reduce complexity and external dependencies, \textsc{SWIFT} reuses the same LLM for  candidate generation and employs key-conditioned selection to embed detectable watermarks while preserving semantic and factual consistency. We next describe these mechanisms and watermark detection.

\noindent \textbf{Instruction-guided Candidate Generation.}
To preserve output utility without auxiliary NLP tools or LMs, \textsc{SWIFT} uses instruction-guided candidate generation to identify watermarkable words and generate substitutions using the same LLM. A key insight is that replacing words with context-aware synonyms can preserve meaning while protecting key information (e.g., named entities or sensitive spans) to ensure factual accuracy.  Accordingly, the instructions guide the LLM to identify protected information,  replaceable words,  and generate  substitutions under semantic and factual constraints (Fig.~\ref{fig:detailed_instruction}). Once admitted by the adaptive scheduler, each completed segment $s_i=(w_{i,1},w_{i,2},\ldots,w_{i,n_i})$, where $w_{i,j}$ is its $j$-th word, is processed through this pipeline. The LLM identifies watermarkable positions $\mathcal{I}_i \subseteq \{1,\ldots,n_i\}$, typically replaceable nouns, verbs, adjectives, and adverbs that can be substituted without altering the  meaning or factual content of $s_i$.


For each segment $s_i$, \textsc{SWIFT} constructs a watermarking   prompt $P_i=[P_{\mathrm{shared}};s_i]$, where $P_{\mathrm{shared}}$ is a shared instruction prefix and the completed segment $s_i$ is the segment-specific suffix. The shared prefix specifies protected information rules, semantic and factual constraints, candidate-generation instructions, and few-shot examples (Appx.~\ref{appd:template}). Since $P_{\mathrm{shared}}$ is identical across watermark requests, vLLM prefix caching reuses its KV states, reducing repeated prefill computation. This is particularly beneficial because the shared prefix is substantially longer than the segment-specific suffix.

Watermark requests are processed via \textit{\textbf{vLLM continuous batching}} \citep{vllm_docs}, joining the shared serving batch as slots become available. This enables dynamic scheduling and concurrent execution of text generation and watermark requests with varying sequence lengths \citep{yu2022orca}.

To ensure consistent candidate generation, \textsc{SWIFT} leverages \textit{\textbf{vLLM structured output decoding}} to constrain responses to a JSON schema.   For each watermarkable
position $j\in\mathcal{I}_i$, let $ \mathcal{C}_{i,j}$ 
denote the set of context-aware candidate substitutions for $w_{i,j}$ (e.g.,  \textit{\{``beautiful'': [``wonderful'', ``scenic'', ``magnificent'']\}}). The structured output for segment $s_i$ is     represented as $\mathcal{C}_i = \left\{ w_{i,j} : \mathcal{C}_{i,j}
\mid j \in \mathcal{I}_i
\right\}$. This reduces malformed outputs and provides   consistent candidate mapping for watermark selection.

\noindent \textbf{Key-conditioned Candidate Selection.}
For each completed segment $s_i$ with its context-aware candidate set $\mathcal{C}_i$, \textsc{SWIFT} embeds the watermark using key-conditioned Tournament selection \citep{dathathri2024scalable}. The selected substitutions are defined as:
$\mathcal{R}_i=\mathbb{T}(k,\mathcal{C}_i,s_i)=\{c^{*}_{i,j}\}_{j\in\mathcal{I}_i}$,
where $\mathbb{T}$ is the Tournament selection function parameterized by secret key $k$, and $c^{*}_{i,j}\in\mathcal{C}_{i,j}$ is the selected candidate for word $w_{i,j}$. Using key-derived pseudorandomness, $\mathbb{T}$ selects candidates reproducibly under the same key but unpredictably without it. The watermarked segment $\tilde{s}_i=\operatorname{Replace}(s_i,\mathcal{R}_i)$ is obtained by applying these substitutions $\mathcal{R}_i$ to $\mathcal{I}_i$ while leaving other words in $s_i$  unchanged. The segment is then streamed as described in Section~\ref{sec:dual_stream}. Since candidate generation enforces semantic, factual, and grammatical constraints, \textsc{SWIFT} avoids additional semantic reranking with external scoring models or similarity filters, reducing model loading, memory overhead, and   complexity.


\noindent \textbf{Watermark Detection.}
\textsc{SWIFT} verifies watermarked text using a key-conditioned contrastive detector, which determines whether a text is watermarked under a specific secret key \citep{dang2026safeseal}. Given a text $y$ and key $k$, the detector predicts $\mathcal{D}(y,k)\in\{0,1\}$, where $\mathcal{D}(y,k)=1$ indicates that $y$ is watermarked under $k$, and $\mathcal{D}(y,k)=0$ otherwise. The detector encodes $y$ with a bidirectional Transformer encoder (RoBERTa \citep{liu2019roberta}), maps $k$ to a deterministic key embedding, and  combines the text and key representations (element-wise interaction and  absolute difference) for classification. 
During training, matched text-key pairs are positives, while mismatched ones are negatives. The detector is optimized with a detection loss for binary classification and a contrastive loss that aligns matched pairs while separating mismatched ones, enabling provider-specific detection. 

\textit{Additional detailed time and space complexity analyses of \textsc{SWIFT} are provided in Appx.~\ref{app:Tech}.}

\section{Experiments}  
Our extensive evaluation covers
\textbf{(1)} watermark effectiveness and efficiency in terms of utility, detectability, and latency compared with representative baselines; \textbf{(2)} robustness against watermark removal attacks; \textbf{(3)} generalization across domains, tasks, and LLMs; \textbf{(4)} adaptive scheduling analysis; \textbf{(5)} key-conditioned watermark detection; and \textbf{(6)} qualitative analysis and ablation studies.

\subsection{Experimental Settings}

\noindent \textbf{Baselines.} We compare against representative LLM watermarks: \textit{KGW} \citep{kirchenbauer2023watermark}, a logit-bias method that steers decoding toward preferred tokens; \textit{SynthID} \citep{dathathri2024scalable}, a sampling-based method using pseudorandom keys; and \textit{SafeSeal} \citep{dang2026safeseal}, a lexical substitution method with named-entity protection. These baselines cover two watermarking categories.

\noindent \textbf{Watermark Removal Attacks.} We evaluate robustness against three attacks: \textit{Dipper}, a sentence-level paraphrasing attack \citep{krishna2024paraphrasing}; \textit{Substitution}, a token-level synonym substitution \citep{pan2024markllm}; and \textit{SIRA}, an LLM-based self-information rewriting attack \citep{cheng2025revealing}.

\noindent \textbf{Tasks, Datasets, and LLMs.}
We evaluate \textsc{SWIFT} across four tasks:   text generation on C4 dataset \citep{dodge2021documentinglargewebtextcorpora},   summarization on CNN/Daily Mail dataset \citep{hermann2015teaching}, mathematical reasoning and short text watermarking on GSM8K dataset \citep{cobbe2021training}, and   technical medical question answering on MedMCQA dataset \citep{pal2022medmcqa}. For C4, we use the first $50$ tokens as prompts and generate up to $200$ tokens.  Other datasets follow their respective summarization or question-answering formats.
 Gemma-3-27B-IT \citep{team2025gemma} serves as the shared LLM, while Qwen2.5-32B \citep{Yang2024Qwen25TR}, LLaMA-3.1-8B-Instruct \citep{grattafiori2024llama}, and Ministral-8B-Instruct \citep{mistral2024ministral8b} support cross-LLM and key-conditioned detection experiments.  We sample $20,000$ examples for detector training and $1,000$ held-out examples for evaluation.

\noindent \textbf{Evaluation Metrics.}
We evaluate \textsc{SWIFT} along four key dimensions:
utility, detectability, robustness, and serving efficiency.
\textbf{\textit{Utility}} measures preservation of semantic
meaning and factual consistency, assessed by GPT-5.5
\citep{openai2026gpt55} as an LLM-as-a-judge on a 0-5 scale.
\textbf{\textit{Detectability}} measures the detector's ability to distinguish
watermarked from non-watermarked outputs using balanced detection accuracy averaged across both output classes.
\textbf{\textit{Robustness}} measures detection accuracy after watermark
removal attacks, reported as the fraction of attacked watermarked
outputs correctly classified as watermarked. \textbf{\textit{Serving efficiency}} is measured primarily by end-to-end latency. 

\textit{Additional metrics and results on serving efficiency, statistical significance, ROC analysis, and scalability, along with  detailed   definitions, calculations, and implementation details, are   in Appx.~\ref{appd:addExpAna}.}


\subsection{Experimental Results and Analysis} \label{sec:main_results}

\begin{figure}[t]
\centering
\subfigure[Text generation]{
\includegraphics[scale=0.2168]{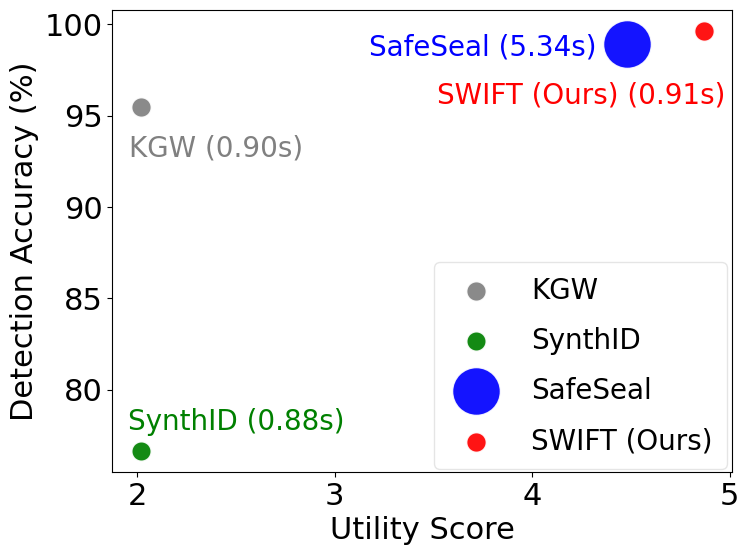}
\label{fig:text_generation_gemma}
}\hfill
\subfigure[Text summarization]{
\includegraphics[scale=0.2168]{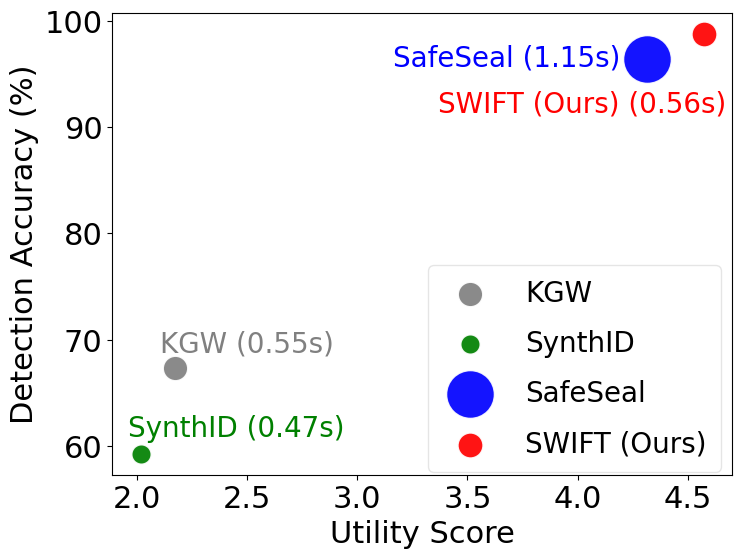}
\label{fig:text_summarization_gemma}
}\hfill
\subfigure[Attack robustness]{
\includegraphics[scale=0.2252]{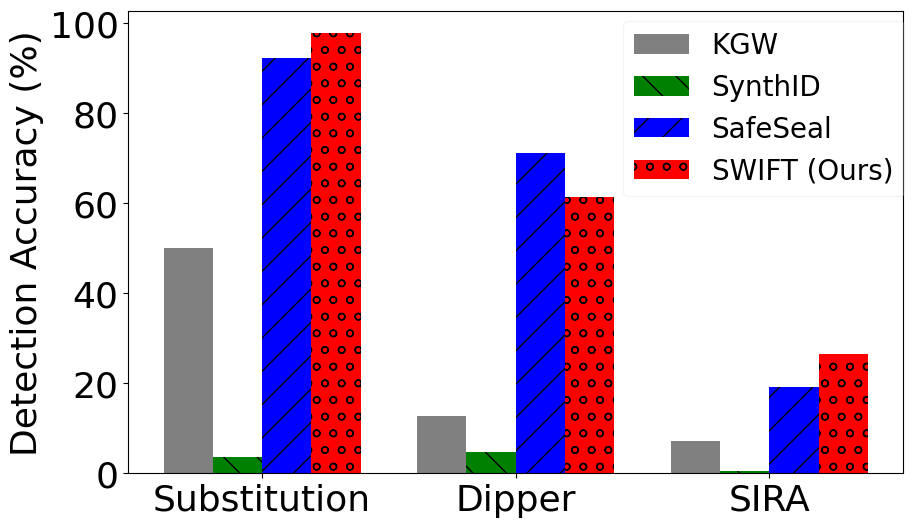}
\label{fig:robustness_gemma}
}
\caption{Watermark effectiveness and efficiency on (a) text generation and (b) summarization, with circle area proportional to latency, and (c) robustness against watermark removal attacks.}
 \vspace{-7.5pt}
\label{fig:LLM_WM}
\end{figure}

\noindent \textbf{Watermark Effectiveness and Efficiency.} 
Fig.~\ref{fig:LLM_WM} demonstrates that \textsc{SWIFT} achieves the best overall trade-off among utility, detectability, and serving latency.  \textbf{\textit{On text generation}}, \textsc{SWIFT} attains the highest utility (4.87) and detection accuracy (99.65\%) with only 0.905s latency. KGW and SynthID are slightly faster (0.88-0.90s) but achieve substantially lower utility (2.01-2.02) and detection accuracy (76.65-95.50\%). SafeSeal achieves strong watermark quality but incurs 5.34s latency, making \textsc{SWIFT} approximately $5.9\times$ faster. \textbf{\textit{On text summarization}} (Fig.~\ref{fig:text_summarization_gemma}), \textsc{SWIFT} similarly achieves the highest utility (4.57) and detection accuracy (98.8\%) with only 0.559s latency. Although KGW and SynthID are slightly faster, their detection accuracies drop to 67.35\% and 59.25\%, respectively. SafeSeal reaches 96.45\% detection but with lower utility and approximately $2\times$ higher latency. Similar trends are observed on Qwen for both text generation and summarization (Fig.~\ref{fig:LLM_WM_qwen}), demonstrating consistent performance across LLMs.
 These results highlight the benefit of co-designing watermarking with inference infrastructure. Instruction-guided candidate generation preserves semantic and factual quality, while key-conditioned selection provides strong detectability. Meanwhile, asynchronous co-serving, adaptive scheduling, and vLLM optimizations reduce watermarking overhead, allowing \textsc{SWIFT} to approach the latency of in-processing methods while substantially improving utility and detectability, and to remain consistently much faster than post-processing SafeSeal.


\noindent \textbf{Robustness Against Attacks.}
Fig.~\ref{fig:robustness_gemma} shows that \textsc{SWIFT} remains highly robust to watermark removal attacks, retaining 97.7\% detection rate under Substitution and 61.3\% under Dipper. \textsc{SWIFT} and SafeSeal substantially outperform KGW and SynthID. This robustness is partly enabled by the key-conditioned detector, which learns watermark patterns from text-key pairs rather than predefined rules or statistical scores that may be disrupted by attacks. SIRA reduces \textsc{SWIFT}'s detection rate to 26.4\%, but at substantial cost. Its attacked outputs achieve only a 0.145 pairwise win rate against the original watermarked outputs using the same LLM judge, indicating considerable utility degradation. SIRA also requires 8.933s, about $9.87\times$ \textsc{SWIFT}'s 0.905s latency. Therefore, while SIRA weakens detectability, it incurs substantial utility loss and runtime overhead.


\noindent \textbf{Domain and Task Generalization.}
Beyond general text generation and summarization, practical watermarking should remain effective on short outputs and technical domains where constrained token choices can reduce watermark capacity. \textit{\textbf{On short mathematical reasoning}} (Table~\ref{tab:gsm8k_watermark}), \textsc{SWIFT} preserves the original task accuracy (without watermark) of 93.2\%, achieves the highest utility (4.93), 90.70\% detection accuracy, and the lowest watermarking latency (0.352s). KGW and SynthID slightly reduce task accuracy and achieve only 53.35\% and 54.95\% detection, respectively. SafeSeal preserves task accuracy with comparable detection (90.75\%) but requires 3.253s, making \textsc{SWIFT} over $9\times$ faster. \textit{\textbf{On highly technical medical question answering}} with constrained watermarking choices (Table~\ref{tab:medmcqa_watermark}), \textsc{SWIFT} maintains the baseline accuracy of 62.6\% without degradation while achieving the highest utility (4.70), 96.70\% detection accuracy, and the lowest latency (0.401s). Together, these results show that \textsc{SWIFT} generalizes across domains and tasks while maintaining strong utility, detectability, and serving efficiency. Its instruction-guided watermarking selectively modifies suitable words under semantic and factual constraints, enabling detectable signals without altering task answers. In contrast, KGW and SynthID make watermarking decisions at every decoding step; limited token choices causes fewer substitutions, weakening watermark strength and detectability. \textsc{SWIFT}'s lower latency on reasoning and medical tasks is further explained by fewer watermark requests and lower watermark density, reducing workload, whereas KGW and SynthID operate at every decoding step. Detailed task characteristics are  in Table~\ref{tab:task_characteristics}.

\begin{table}[t]
\centering
\small
\caption{Performance on GSM8K. $\Delta$Acc. is the relative accuracy change from the No watermark.}
\label{tab:gsm8k_watermark}

\begin{tabular}{lccccc}
\toprule
\textbf{Setting} 
& \textbf{Accuracy (\%) $\uparrow$} 
& \textbf{$\Delta$Acc. (\%) $\uparrow$}
& \textbf{Utility $\uparrow$} 
& \textbf{Detection (\%) $\uparrow$} 
& \textbf{Latency (s) $\downarrow$} \\
\midrule
No watermark 
    & 93.2 
    & -- 
    & --
    & -- 
    & 0.208 \\

KGW 
    & 92.3 
    & -0.97
    & 4.66
    & 53.35 
    & 0.445 \\

SynthID 
    & 92.7 
    & -0.54
    & 4.45
    & 54.95 
    & 0.437 \\

SafeSeal 
    & \textbf{93.2} 
    & \textbf{0.00}
    & 4.61
    & \textbf{90.75} 
    & 3.253 \\

\textsc{SWIFT} (Ours)
    & \textbf{93.2} 
    & \textbf{0.00}
    & \textbf{4.93}
    & 90.70 
    & \textbf{0.352} \\
\bottomrule
\end{tabular}
\end{table}

\begin{figure}[t]
\centering
\subfigure[Scheduler Latency Performance]{
\includegraphics[scale=0.25]{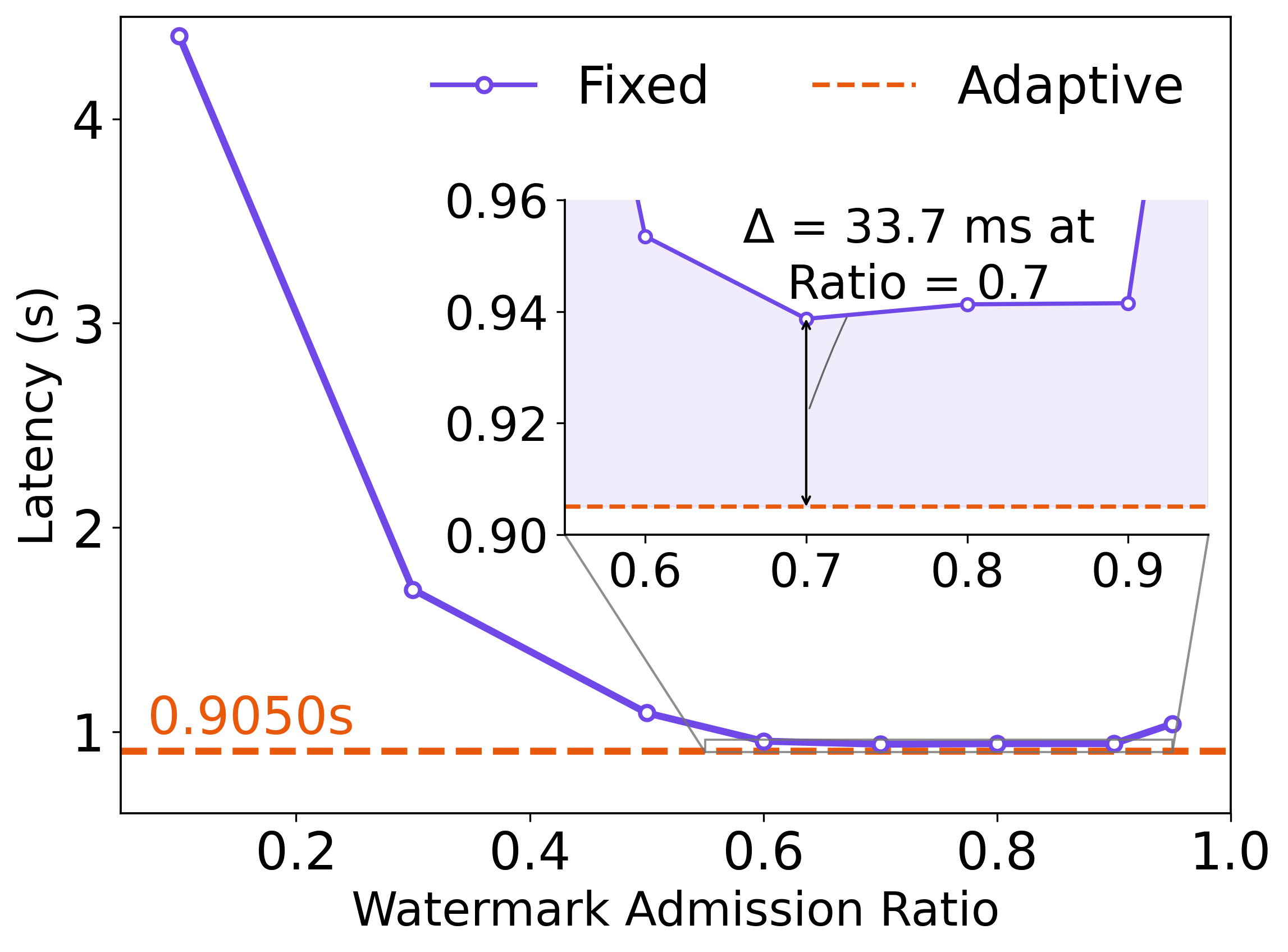}
\label{fig:adaptive_scheduler}
}\hspace{2cm}
\subfigure[Key-conditioned Detection]{
\includegraphics[scale=0.26]{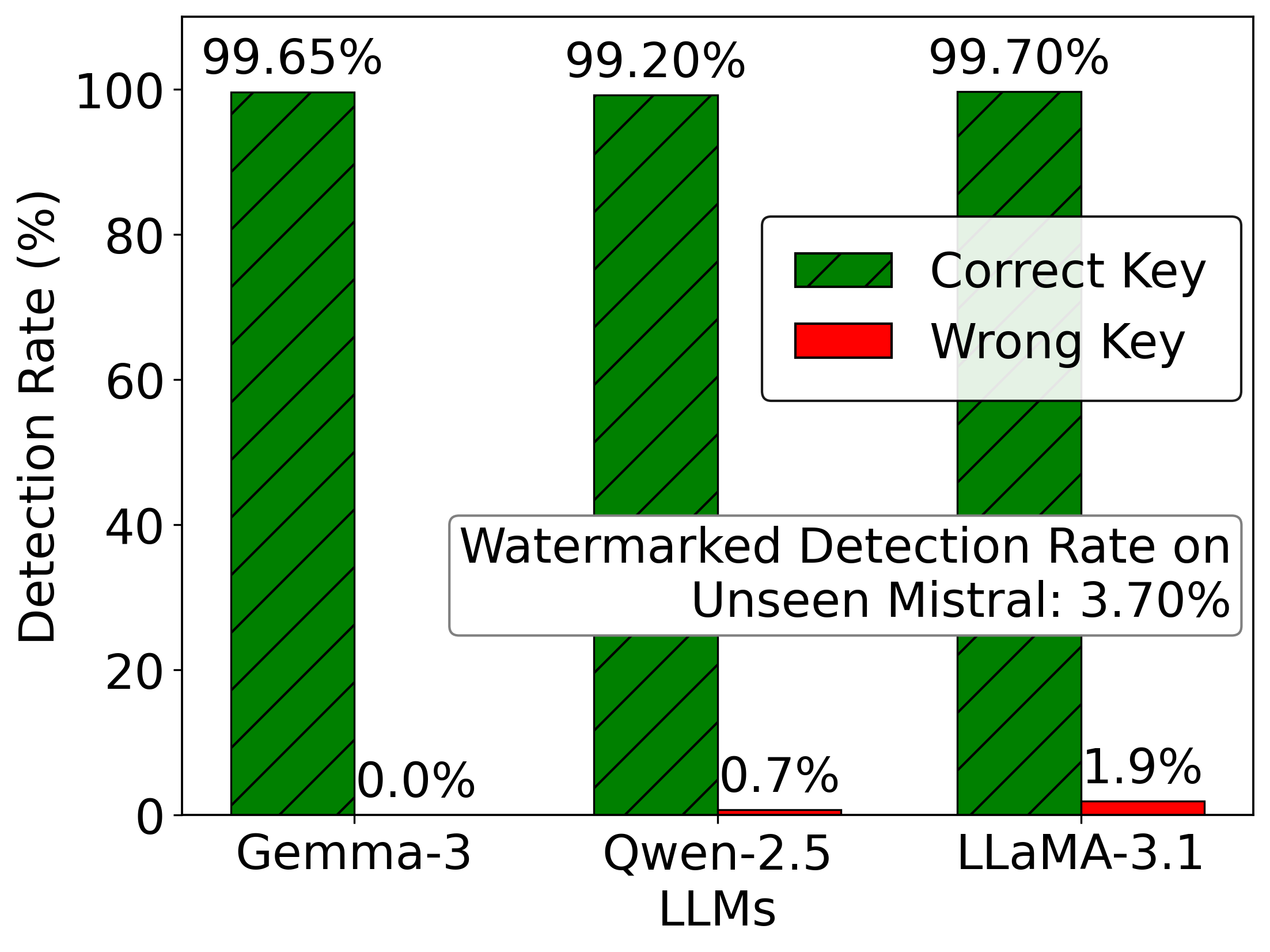}
\label{fig:key_detection}
}
\caption{Core serving evaluation and key-conditioned detection performance.}
\vspace{-10pt}
\label{fig:Serving_metrics}
\end{figure}

\noindent \textbf{Adaptive Co-serving Scheduler.}
Fig.~\ref{fig:adaptive_scheduler} compares \textsc{SWIFT}'s adaptive scheduler with fixed admission ratios in $[0.1,0.95]$. The adaptive scheduler updates every five scheduling steps and achieves the lowest   latency of 0.905s.  Low fixed ratios allocate insufficient capacity to watermarking, causing request backlogs and higher latency, while an overly high ratio ($r=0.95$) interferes with text generation and   increases latency. In contrast, \textsc{SWIFT} adapts watermark admission to runtime workload pressure, balancing both streams while maintaining low latency and streaming responsiveness.


\noindent \textbf{Key-conditioned Detection.}
Fig.~\ref{fig:key_detection} evaluates \textsc{SWIFT}'s universal detector across LLMs using correct and incorrect keys. With the correct key, detection remains high at 99.2-99.7\% across Gemma-3, Qwen-2.5, and LLaMA-3.1, while incorrect-key detection remains near zero (0-1.9\%), demonstrating strong key specificity and reducing false attribution. The key-conditioned objective aligns matched text-key pairs while separating mismatched pairs, enabling key-specific verification. On the unseen Ministral model, only 3.7\% of outputs are detected using the Gemma-3 key, further indicating that the detector does not associate unseen models with known keys. A single detector can therefore generalize across LLMs without per-model retraining, simplifying system integration.


\begin{table*}[t!]
\small
\centering
\caption{Text utility comparison of original and watermarked outputs. Green marks \hl{factual consistency}, blue marks \hlb{meaning-preserving substitutions}, and red marks \hlred{semantic or factual errors}.}
\begin{adjustbox}{width=1.0\textwidth}
\begin{tabular}{|p{0.097\linewidth}|p{1.03\linewidth}|}
\hline

\centering \textbf{Prompt} &
\raggedright
 \ldots Chief inspector Neil Kentish joined Thames Valley Police less than a month ago after serving in \ldots
\tabularnewline \hline

\centering
\multirow{4}{*}{\makecell{\textbf{No} \\ \textbf{watermark}}} &
\raggedright
the Metropolitan Police for 20 years. He's taken over from chief inspector Olly Wright, who has moved to a new role within the force. Chief Insp Kentish said he's been spending his time getting to know the area, meeting local officers and community leaders.
``It's been a really busy few weeks, but a really positive few weeks'' he said. ``Aylesbury is a really interesting place, it’s got a lot going for it.'' ...
\tabularnewline \hline

\centering
\multirow{5}{*}{\makecell{\textbf{KGW}}} &
\raggedright
the \hlb{Met} \hl{Police} for \hlred{25 years}.
He heads up the Wycombe and \hl{Aylesbury} areas \hlred{as deputy area commander}, reporting \hlred{to area commander for Buckinghamshire and Milton Keynes, Chief Superintendent Rob Harding.}

\hl{Chief Insp Kentish} said:
\hlred{``It’s very different coming from London}. \hlb{Aylesbury is a lovely market town}, \hlred{the pace of life is noticeably slower but people are really friendly''.} 
\hlb{``I've been out and about meeting community groups,''} ...
\tabularnewline \hline

\centering
\multirow{4}{*}{\makecell{\textbf{SynthID}}} &
\raggedright
\hl{Metropolitan Police} for \hlred{23 years.}
He is taking over from chief \hlred{inspector Rick Carr, who recently transferred to the Roads Policing Unit.}
\hl{Chief Inspector Kentish} \hlb{has made it a priority to
meet local communities
and businesses and key partners.}
\hlred{``I’ve been really impressed with the commitment and dedication of the Aylesbury local policing area team. They are really enthusiastic and want to do the best for the community.''} ...
\tabularnewline \hline

\centering
\multirow{5}{*}{\makecell{\textbf{SafeSeal}}} &
\raggedright
\hl{the Metropolitan Police} for \hl{20 years.}
He’s taken over from
\hlred{chief inspection} \hl{Olly Wright},
who has \hlb{bolted} to a \hlb{new work}
within the \hlb{commission}.
\hl{Chief Insp Kentish} \hlb{explained} he’s been
\hlb{spent his work starting to know the community},
\hlb{meeting local commanders}
and \hlb{community representatives}. ``It's been a
\hlb{pretty busy} few weeks, but a
\hlb{totally positive} few weeks,'' he \hlb{added.} \hl{``Aylesbury} is a
\hlb{seriously interesting town},
it’s got a lot \hlred{riding} for it.'' ...
\tabularnewline \hline

\centering
\multirow{4}{*}{\makecell{\textbf{\textsc{SWIFT}} \\ \textbf{(Ours)}}} &
\raggedright
\hl{the Metropolitan Police} for \hl{20 years.}
He’s \hlb{stepped in over} from
\hl{chief inspector Olly Wright},
who has \hlb{transferred to another role}
within the force.
\hl{Chief Insp Kentish} said he’s been
\hlb{devoting his time learning to know the area},
\hlb{encountering} local officers
and community leaders.
``It’s been a really
\hlb{hectic} few weeks, but a really
\hlb{nice} few weeks'' he said.
\hl{``Aylesbury} is a really
\hlb{engaging place},
it’s got \hlb{a lot happening for it}.'' ...
\tabularnewline \hline
\end{tabular}
\end{adjustbox}
\label{table:summarization_example}
\vspace{-10pt}
\end{table*}

\noindent \textbf{Qualitative Evaluation.}
To further examine text quality, Table~\ref{table:summarization_example} compares the original response with outputs from different watermarking methods. \textsc{SWIFT} best preserves   factual consistency with mostly meaning-preserving substitutions, whereas KGW and SynthID exhibit substantial factual drift and SafeSeal introduces less appropriate substitutions. KGW changes the duration from 20 to 25 years and adds unsupported roles and personnel. SynthID changes it to 23 years  and adds unsupported policing details. These examples show how token-level perturbations during autoregressive decoding can propagate into broader   deviations. SafeSeal preserves the main facts but changes ``a lot going for it'' to ``a lot riding for it.'' 
In contrast, \textsc{SWIFT} preserves key facts while making context-aware changes such as ``a lot happening for it.'' By generating context-aware candidates with the shared LLM, \textsc{SWIFT} eliminates external dependencies while improving semantic control and limiting factual drift. Additional reasoning and medical examples (Appx.~\ref{appd:addExpAna}) show similar trends.

\begin{table}[t]
\centering
\footnotesize
\caption{Ablation Study: Comparison of \textsc{SWIFT} Variants.}
\label{tab:swift_ablation}

\setlength{\tabcolsep}{5pt}
\begin{tabular}{@{}lcccc@{}}
\toprule
\textbf{Version}
& \textbf{Latency (s) $\downarrow$}
& \textbf{$\Delta$ Latency (\%) $\downarrow$}
& \textbf{Utility $\uparrow$}
& \textbf{$\Delta$ Utility (\%) $\uparrow$} \\
\midrule
\textsc{SWIFT}
    & 0.905
    & --
    & 4.871
    & -- \\

SWIFT-FB
    & 1.0921
    & +20.67
    & 4.871
    & 0.00 \\

SWIFT-NER
    & 1.7005
    & +87.90
    & 4.610
    & -5.36\\

SWIFT-EN
    & 0.900
    & -0.55
    & 2.389
    & -50.95 \\

\bottomrule
\end{tabular}
\vspace{-10pt}
\end{table}

\noindent \textbf{Ablation Studies.}
Table~\ref{tab:swift_ablation} evaluates key \textsc{SWIFT} components. SWIFT-FB disables continuous batching and request prioritization, increasing latency by 20.67\% without affecting utility. SWIFT-NER replaces LLM-based entity handling with an external NER tool (spaCy \citep{spaCy2020}), increasing latency by 87.90\% and reducing utility by 5.36\%. SWIFT-EN removes entity-protection instructions, improving latency by 0.55\% while reducing utility by 50.95\%. Overall, serving optimizations, a shared LLM, and entity protection jointly improve \textsc{SWIFT}'s latency-utility trade-off.

\textbf{Limitations.} Despite its effectiveness, \textsc{SWIFT} has several limitations. First, \textit{watermark quality depends on the capability of the shared LLM}. Since the same model performs text generation, watermarkable-word identification, and candidate substitution, weaker models may produce less reliable substitutions or fail to preserve protected information, reducing output quality and utility in specialized domains. Second, \textsc{SWIFT} \textit{does not fully leverage KV-cache reuse}. While shared instructions benefit from prefix caching, reusing KV states for segment-specific suffixes remains non-trivial because text generation uses the original prompt and history, while candidate generation uses a separate watermarking context. Finally, although \textsc{SWIFT} achieves competitive serving efficiency (Table~\ref{tab:ttft_tpot}), it \textit{currently streams outputs at the sentence level} rather than token by token.


\section{Conclusion}

We presented \textsc{SWIFT}, a serving-aware LLM watermarking framework that co-designs watermark construction with modern inference infrastructure. \textsc{SWIFT} unifies text generation and watermark construction within a shared LLM backend, reducing reliance on auxiliary models and external NLP tools, while instruction-guided candidate generation and entity protection preserve semantic and factual utility. Its asynchronous co-serving architecture and adaptive scheduler coordinate text generation and watermark workloads under varying demand while leveraging inference-engine optimizations. Experiments across domains, tasks, and LLMs demonstrate strong utility, detectability, robustness, downstream task preservation, and serving efficiency. These results highlight the benefits of jointly considering watermark quality and inference-system design. Future work will explore finer-grained streaming through adaptive token release and improved KV-cache reuse across generation and watermarking contexts, further improving responsiveness and computational efficiency.

\bibliography{SWIFT}
\bibliographystyle{iclr2027_conference}

\newpage
\appendix
\section{Additional Technical Details and Analysis} \label{app:Tech}


\textbf{Related Work.} 
Table~\ref{tab:Summary} compares \textsc{SWIFT} with existing watermarking methods across key characteristics relevant to practical implementation. In-processing methods naturally support streaming but generally lack explicit named-entity preservation, while post-processing methods can protect entities but typically require complete outputs and auxiliary LMs or NLP tools for watermark construction. Most existing methods are developed using frameworks such as Hugging Face Transformers, with limited integration of   serving optimizations. In contrast, \textsc{SWIFT} combines named-entity preservation, streaming generation, and watermark construction without auxiliary models or NLP tools within a shared vLLM backend, enabling efficient watermarking through optimized inference and asynchronous co-serving.

\begin{table}[H]
\small
    \centering
    \caption{Comparison of \textsc{SWIFT} with representative techniques across key characteristics.}
    \renewcommand{\arraystretch}{1.0}
   \begin{adjustbox}{width=0.8\textwidth}
    \begin{tabular}{p{4.3cm}  p{2.3cm} >{\centering\arraybackslash}m{1.7cm} >{\centering\arraybackslash}m{1.6cm} >{\centering\arraybackslash}m{1.5cm} >{\centering\arraybackslash}m{2cm}}
\toprule
\textbf{Method} 
& \textbf{Generation} 
& \makecell{\textbf{Named Entity}\\ \textbf{Preservation}} 
& \makecell{ \textbf{Streaming}\\\textbf{Generation}} 
& \makecell{\textbf{No Auxiliary} \\\textbf{Models}}
& \makecell{\textbf{Serving}\\\textbf{System}}\\
\midrule
       KGW \citep{kirchenbauer2023watermark}  & In-processing    & \ding{55}  & \ding{51} & \ding{51} & Hugging Face \\
       SynthID \citep{dathathri2024scalable}  & In-processing    & \ding{55}  & \ding{51} & \ding{51} & Hugging Face\\
       EXP \citep{kuditipudi2023robust}  & In-processing    & \ding{55}  & \ding{51} & \ding{51} & Hugging Face\\
       SIR \citep{liu2024semanticinvariantrobustwatermark}  & In-processing    & \ding{55}  & \ding{51} & \ding{55} & Hugging Face\\
       SemaMark \citep{ren2023robust} & In-processing    & \ding{55}  & \ding{51} & \ding{55} & Hugging Face\\
       SemStamp  \citep{hou2023semstamp} & In-processing    & \ding{55}  & \ding{51} & \ding{55} & Hugging Face\\
        DAWA  \citep{he2026theoretically} & In-processing    & \ding{55}  & \ding{51} & \ding{51} & Hugging Face\\
       CATER \citep{he2022cater}  & Post-processing & \ding{51}  & \ding{55} & \ding{51}  & Hugging Face\\
       TW \citep{yang2023watermarking} & Post-processing & \ding{51}  & \ding{55} & \ding{55}  & Hugging Face\\
       DTM \citep{munyer2024deeptextmark} & Post-processing & \ding{51}  & \ding{55} & \ding{55}  & Hugging Face\\
       PostMark  \citep{chang2024postmark}  & Post-processing & \ding{51}  & \ding{55} & \ding{55}  & Hugging Face\\
      SafeSeal \citep{dang2026safeseal} & Post-processing & \ding{51}  & \ding{55} & \ding{55}  & Hugging Face\\
        \midrule
        \textbf{\textsc{SWIFT} (Ours) } & Post-processing & \ding{51} & \ding{51} & \ding{51}  & vLLM\\ 
        \bottomrule
    \end{tabular}
    \end{adjustbox}
    \label{tab:Summary}
\end{table}

 \textbf{Algorithm.}  Algorithm~\ref{alg:swift} summarizes the end-to-end \textsc{SWIFT} pipeline,
including asynchronous generation, segment buffering, watermark insertion,
and key-conditioned detection.

\begin{algorithm}[H]
\small
\caption{\textsc{SWIFT} Algorithm}
\label{alg:swift}
\begin{algorithmic}[1]

\STATE \textbf{Inputs}: user prompts $\mathcal{X}=\{x_1,\ldots,x_N\}$, 
LLM $\theta$, secret key $k$, queue length threshold $B_{\max}$, waiting time threshold $W_{\max}$, admission bounds $r_{\min},r_{\max}$, scheduling steps $K$, watermark queue $\mathcal{Q}^{wm}$, batch size $B$, shared instruction $P_{\mathrm{shared}}$

\STATE \textbf{Outputs}: streaming watermarked text 
$\{y_1^{wm},\ldots,y_N^{wm}\}$ and detector prediction $\mathcal{D}(y,k)$

\STATE \textbf{Asynchronous Text Generation}:
\begin{ALC@g}
\FOR{each user prompt $x_u \in \mathcal{X}$}
    \STATE Generate async text via vLLM: $y_u \leftarrow \theta(x_u)$
    \FOR{each streamed token $y_{u,t}$ from $y_u$}
        \STATE Append $y_{u,t}$ to the partial output buffer for user $u$
        \IF{a complete sentence segment $s_i$ is identified}
            \STATE Add $s_i$ to the watermark queue $\mathcal{Q}^{wm}$
        \ENDIF
    \ENDFOR
\ENDFOR
\end{ALC@g}

\STATE \textbf{Adaptive Co-Serving Scheduler}:
\begin{ALC@g}
\STATE At scheduling step $\tau$, obtain watermark queue length 
$q_\tau = |\mathcal{Q}^{wm}_\tau|$ and oldest-request waiting time $w_\tau$
\STATE Every $K$ scheduling steps, update pressure $p_\tau$ and watermark admission ratio $r_\tau$
\STATE Compute watermark capacity 
$B_\tau^{wm}=\lfloor r_\tau B \rfloor$ and admit  $B_\tau^{wm}$ requests to the shared serving batch
\end{ALC@g}

\STATE \textbf{Asynchronous LLM Watermarking}:
\begin{ALC@g}
\WHILE{$\mathcal{Q}^{wm}$ is not empty}
    \STATE Collect admitted completed segments:
    $\mathcal{S}_\tau=\{s_1,\ldots,s_m\}$

    \STATE Form instruction-guided candidate-generation prompts:
    $P_i=[P_{\mathrm{shared}};s_i], \quad s_i\in\mathcal{S}_\tau$
    \STATE Identify positions $\mathcal{I}_i$ and generate  candidates:
        $\mathcal{C}_i =
        \{w_{i,j}:\{c_{i,j,1},\ldots,c_{i,j,M_{i,j}}\}
        \mid j\in\mathcal{I}_i\}$

    \FOR{each sentence segment $s_i\in\mathcal{S}_\tau$}
        \STATE Select key-conditioned substitutions:
            $\mathcal{R}_i=\mathcal{T}(k,\mathcal{C}_i,s_i)
            =\{c^{*}_{i,j}\mid j\in\mathcal{I}_i\}$
        \STATE Apply $\mathcal{R}_i$ to obtain
        $\tilde{s}_i=\operatorname{Replace}(s_i,\mathcal{R}_i)$
        \STATE Emit the watermarked segment $\tilde{s}_i$
    \ENDFOR
\ENDWHILE

\STATE \textbf{Return}: streaming watermarked outputs 
$\{y_1^{wm},\ldots,y_N^{wm}\}$
\end{ALC@g}

\STATE \textbf{Watermark Detection}:
\begin{ALC@g}
\STATE Given text $y$ and secret key $k$, apply the key-conditioned detector:
   $\mathcal{D}(y,k)\in\{0,1\}$
\STATE where $\mathcal{D}(y,k)=1$ indicates that $y$ is watermarked under $k$, 
and $\mathcal{D}(y,k)=0$ otherwise
\STATE \textbf{Return}: detector prediction $\mathcal{D}(y,k)$
\end{ALC@g}

\end{algorithmic}
\end{algorithm}

\begin{figure*}[!htbp]
      \centering    
      \includegraphics[scale=0.068]{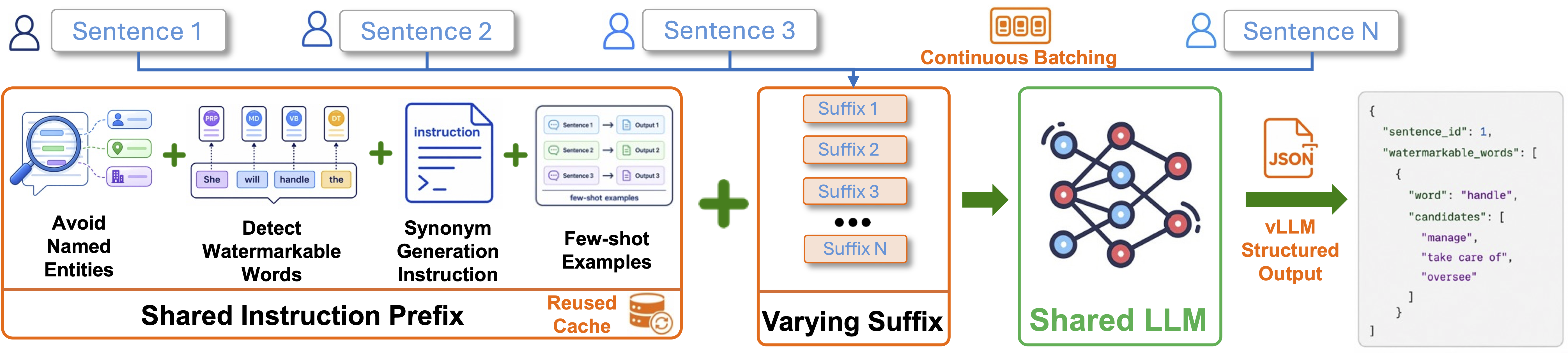} 
      \caption{Detailed Instruction-guided Candidate Generation with optimization techniques.}
      \label{fig:detailed_instruction}
 \end{figure*}

\textbf{Instruction-guided Candidate Generation.}  
 Fig.~\ref{fig:detailed_instruction} illustrates the instruction-guided candidate generation process for the watermark construction stream. Each watermark request combines a reusable shared instruction prefix - covering named-entity protection, watermarkable-word detection, synonym-generation constraints, and few-shot examples -with a sentence-specific suffix. The shared prefix enables prefix-cache reuse across continuously batched requests, while the unified LLM generates candidates under vLLM structured decoding to produce a consistent JSON mapping from watermarkable words to context-aware substitutions. In addition, Fig.~\ref{fig:instruction_prompt} provides the detailed template of the shared instruction prefix used in \textsc{SWIFT} for candidate generation.

\textbf{System-level Optimizations.} 
Beyond the scheduling mechanism and serving techniques, \textsc{SWIFT} incorporates additional system-level optimizations for efficient shared inference. Both text generation and watermark construction use the same LLM with a \textbf{unified precision} \texttt{bf16} configuration, avoiding duplicate model weights, separate runtimes, and additional memory overhead. \textsc{SWIFT} also leverages \textbf{chunked prefill} \citep{vllm_docs}, which interleaves long watermark-prefill computation with ongoing decoding to reduce interference with latency-sensitive generation \citep{agrawal2023sarathi}. Together, these optimizations improve resource efficiency under concurrent workloads.

\textbf{Time Complexity Analysis.} Let $P$ and $T$ denote the prompt and output lengths, respectively, and let
$N_s$ denote the number of completed segments, where $|s_i|$ is the length
of segment $s_i$. Let $G_i$ denote the number of output tokens generated
for the structured candidate mapping of segment $s_i$. 

\textsc{SWIFT} has two main computational costs: text generation and
watermarking. The generation cost is
$
\mathcal{C}_g(P,T)
=
\mathcal{C}_{\mathrm{prefill}}(P)
+
\sum_{t=1}^{T}
\mathcal{C}_{\mathrm{decode}}(P+t)
$,
including prompt prefill and autoregressive decoding of $T$ tokens.
Maintaining the segment buffer and detecting segment boundaries adds
$\mathcal{O}(T)$ lightweight overhead. Thus, 
$\mathcal{C}_{\mathrm{gen}}
=
\mathcal{O}\left(
\mathcal{C}_g(P,T)+T
\right).$

For watermarking, each segment $s_i$ is processed using
$P_i=[P_{\mathrm{shared}};s_i]$. Without prefix caching, candidate
generation repeatedly processes the shared instruction and generates
the structured candidate mapping:
$
\sum_{i=1}^{N_s}
\left[
\mathcal{C}_{\mathrm{prefill}}
\left(
|P_{\mathrm{shared}}|+|s_i|
\right)
+
\sum_{g=1}^{G_i}
\mathcal{C}_{\mathrm{decode}}
\left(
|P_{\mathrm{shared}}|+|s_i|+g
\right)
\right]
$.

With prefix caching, the shared prefix is computed once and reused across
watermark requests, reducing repeated prefill computation. If
$\mathcal{I}_i$ denotes the watermarkable positions in $s_i$, each with at
most $M$ candidates, key-conditioned selection adds
$
\mathcal{O}\left(
\sum_{i=1}^{N_s}
|\mathcal{I}_i|M
\right).
$
Therefore, the watermarking complexity can be expressed as
\begin{equation}
\small
\begin{aligned}
\mathcal{C}_{\mathrm{wm}}
=
\mathcal{O}\left(
\mathcal{C}_{\mathrm{prefill}}(|P_{\mathrm{shared}}|)
+
\sum_{i=1}^{N_s}
\mathcal{C}_{\mathrm{prefill}}(|s_i|)
+
\sum_{i=1}^{N_s}
\sum_{g=1}^{G_i}
\mathcal{C}_{\mathrm{decode}}
\left(
|P_{\mathrm{shared}}|+|s_i|+g
\right)
+
\sum_{i=1}^{N_s}
|\mathcal{I}_i|M
\right).
\end{aligned}
\end{equation}

The adaptive scheduler introduces only constant-time queue-pressure and
admission computations at each update and therefore does not change the
asymptotic complexity. Overall computational work remains
$\mathcal{O}(\mathcal{C}_{\mathrm{gen}}+\mathcal{C}_{\mathrm{wm}})$.
However, generation and watermark requests are asynchronously co-served
within the same inference engine, allowing their execution to overlap
through mixed continuous batching. Consequently, observed wall-clock
latency depends on the adaptive admission ratio and runtime workload rather
than following the strictly sequential
$\mathcal{C}_{\mathrm{gen}}+\mathcal{C}_{\mathrm{wm}}$ execution of
generate-then-watermark pipelines~\citep{dang2026safeseal}.

\textbf{Space Complexity.}
\textsc{SWIFT} uses one unified LLM and shared inference engine for both
generation and watermarking. Its dominant GPU memory cost is therefore
$
\mathcal{O}
\left(
M_{\mathrm{LLM}} + M_{\mathrm{KV}}
\right),
$
where $M_{\mathrm{LLM}}$ denotes model-weight memory and
$M_{\mathrm{KV}}$ denotes the shared KV cache for active generation and
watermark requests, including cached prefix states.

Additional state consists of partial generation segments, the global
watermark queue, and active candidate mappings. Let
$\mathcal{A}_{\mathrm{gen}}$ and $\mathcal{A}_{\mathrm{wm}}$ denote the
active generation and watermark requests, respectively. The total space
complexity is
\begin{equation}
\small
\mathcal{O}\left(
M_{\mathrm{LLM}}
+
M_{\mathrm{KV}}
+
\sum_{u\in\mathcal{A}_{\mathrm{gen}}} |\mathcal{B}_u|
+
\sum_{s_i\in\mathcal{Q}^{\mathrm{wm}}} |s_i|
+
\sum_{i\in\mathcal{A}_{\mathrm{wm}}} |\mathcal{I}_i|M
\right),
\label{eq:space_complexity}
\end{equation}
where $\mathcal{B}_u$ is the partial segment buffer for generation request
$u$ and $\mathcal{Q}^{\mathrm{wm}}$ is the global watermark queue.
Candidate mappings are released after key-conditioned selection. Unlike
pipelines that require separate auxiliary models or watermarking engines,
\textsc{SWIFT} maintains only one copy of the LLM weights on the shared GPU.


\section{Additional Experimental Details and Results}
\label{appd:addExpAna}

\textbf{Implementation Backends and Hardware.}
All experiments use a single NVIDIA A100 80GB GPU, batch size 32, and KV-cache block size 16. For fair comparison, all baselines use the same vLLM backend, model, and generation settings. KGW \citep{kirchenbauer2023watermark} uses vLLM's logits-processing interface, while SynthID \citep{dathathri2024scalable} adapts the vLLM sampler following vLLM-Watermark \citep{vllm_watermark}. For SafeSeal \citep{dang2026safeseal}, text is generated with vLLM before applying its post-processing watermarking procedure. This minimizes backend differences and enables consistent comparison of watermarking overhead. For \textsc{SWIFT}, the adaptive scheduler updates the watermark admission ratio every $K=5$ scheduling steps, with $r_{\min}=0$, $r_{\max}=1$, queue threshold $B_{\max}=32$, and oldest-request waiting time threshold $W_{\max}=1.5$s. These settings are fixed unless otherwise noted.

\textbf{Evaluation Metrics.} We use diverse evaluation metrics to assess watermark utility, detectability, robustness, and serving efficiency, covering both output quality and system-level performance.

\textbf{\textit{Utility.}}
Utility measures preservation of meaning and factual content. We use GPT-5.5 \citep{openai2026gpt55} as an LLM-as-a-judge $J(\cdot)$, assigning each original output $y$ and watermarked output $y^{wm}$ a score from 0-5, with higher scores indicating better preservation. The same judge performs pairwise attack analysis by comparing attacked outputs with their original watermarked versions and reporting the attacked outputs' win rate. The mean utility score over $N$ samples is:
\begin{equation}
\small
    \text{UtilityScore} =
    \frac{1}{N}
    \sum_{n=1}^{N}
    J(y_n, y_n^{wm}).
    \label{eq:utility_score}
\end{equation}

\textbf{\textit{Detectability.}}
Detectability measures whether a detector can distinguish watermarked outputs from non-watermarked outputs. To account equally for both classes, we define detection accuracy as the average of the detector's accuracy on watermarked ($y_i^{wm}$) and non-watermarked ($y_j$) samples:
\begin{equation}
\small
   \text{DetectionAccuracy} =
    \frac{1}{2}
    \left[
    \frac{1}{N}
    \sum_{i=1}^{N}
    \mathbb{I}\left(\mathcal{D}(y_i^{wm},k)=1\right)
    +
    \frac{1}{H}
    \sum_{j=1}^{H}
    \mathbb{I}\left(\mathcal{D}(y_j,k)=0\right)
    \right],
    \label{eq:detection_accuracy}
\end{equation}
where $\mathcal{D}(\cdot)$ is the detector with key $k$, $\mathbb{I}(\cdot)$ is the indicator function, $N$ and $H$ denote the numbers of watermarked and non-watermarked samples, respectively. Higher detection accuracy is better.

\textbf{\textit{Robustness.}}
Robustness evaluates whether a watermark remains detectable after its output is modified by an attack. For an attack $\mathcal{A}$, we compute the post-attack detection accuracy as
\begin{equation}
\small
    \text{DetectionAccuracy}_{\mathcal{A}} =
    \frac{1}{N}
    \sum_{i=1}^{N}
    \mathbb{I}
    \left(
    \mathcal{D}(\mathcal{A}(y_i^{wm}),k) = 1
    \right),
    \label{eq:attack_detection}
\end{equation}
where $\mathcal{A}(y_i^{wm})$ is the attacked watermarked output. Higher accuracy indicates greater robustness.

\textbf{\textit{Latency.}}
We report amortized end-to-end serving time per request, computed as the total
wall-clock time to process the serving workload divided by the number of requests.
This measure accounts for concurrent batched execution and includes text generation,
watermark construction, and output emission. Lower values indicate higher serving
efficiency:
\begin{equation}
\small
L = \frac{t^{\mathrm{end}} - t^{\mathrm{begin}}}{N_{\mathrm{req}}},
\label{eq:latency}
\end{equation}
where $t^{\mathrm{begin}}$ and $t^{\mathrm{end}}$ denote the start and completion
times of the serving workload, respectively, and $N_{\mathrm{req}}$ is the number
of requests.


\textbf{\textit{Throughput.}}
Throughput measures the number of generated output tokens processed per second over the serving workload, with higher throughput indicating better serving efficiency:
\begin{equation}
\small
TP =
\frac{\sum_{n=1}^{N_{\mathrm{req}}} T_n}
{t^{\mathrm{end}}-t^{\mathrm{begin}}},
\label{eq:throughput}
\end{equation}
where $T_n$ is the number of generated tokens for request $n$. 


\textbf{\textit{Peak GPU Memory.}}
Peak GPU memory measures the maximum GPU memory occupied during generation and watermarking. Lower peak memory indicates a smaller deployment footprint.

\textbf{\textit{GPU KV-Cache Usage.}}
GPU KV-cache usage measures the fraction of the engine's KV-cache capacity occupied by active requests. Unlike peak GPU memory, it reflects only the KV-cache pool. Lower usage indicates greater headroom for additional concurrent requests.

\textbf{\textit{Prefix-Cache Hit Rate.}}
Prefix-cache hit rate measures the fraction of reusable prompt-prefix computation served from the prefix cache rather than recomputed during prefill. A higher hit rate indicates greater reuse of previously computed prefix states and reduced redundant computation.

\textbf{\textit{Service TTFT.}}
Service time-to-first-token (TTFT) measures the time from request admission to the first output token, excluding pre-admission waiting. Lower Service TTFT is preferred.

\textbf{\textit{E2E TTFT.}}
End-to-end time-to-first-token (E2E TTFT) measures the time from request submission to the first output token. Unlike Service TTFT, it includes queueing and scheduling delays, providing a workload-level measure of responsiveness. Lower E2E TTFT is preferred.

\textbf{\textit{TPOT.}}
Time per output token (TPOT) measures the average time required to produce each output token after the first token, with lower TPOT indicating faster autoregressive decoding:
\begin{equation}
\small
    TPOT =
    \frac{t^{\mathrm{final}}-t^{\mathrm{first}}}
    {T-1},
    \label{eq:tpot}
\end{equation}
where $t^{\mathrm{first}}$ and $t^{\mathrm{final}}$ denote the timestamps of the first and final output tokens, respectively, and $T$ is the number of generated tokens.


\textbf{Additional Evaluation Setup.}
Unless otherwise specified, all results are computed over $1,000$  samples. In the C4 text generation, we use the first 50 tokens of each C4 sample as the prompt and generate up to 200 output tokens. In the CNN/Daily Mail Summarization, we use a fixed few-shot summarization prompt and instruct the model to generate a summary of at most 60 words. In the GSM8K reasoning task, the model is prompted to generate a reasoning trace of at most 50 tokens followed by a final answer. Task accuracy is computed by comparing the extracted final answer with the ground-truth answer. In the MedMCQA, we use the same reasoning-and-final-answer format as GSM8K but allow explanations of up to 100 tokens. We retain only single-answer questions with a valid correct option and a non-empty explanation longer than 50 words to ensure sufficient text for watermarking and evaluation.

For watermark detection, we train the RoBERTa-large (356M) \citep{liu2019roberta} key-conditioned detector once on $20,000$ C4 samples and evaluate it on $1,000$ held-out samples. Training and evaluation prompts come from the official C4 training and validation splits, ensuring no overlap. The same detector is reused without retraining for summarization, mathematical reasoning, and medical QA, enabling cross-domain and cross-task evaluation under distribution shift.

For utility evaluation, we use GPT-5.5 \citep{openai2026gpt55} as an LLM-as-a-judge to evaluate semantic and factual preservation between the original and watermarked outputs using the fixed prompt shown in Fig.~\ref{template:llm-judge}. The reported utility is the average score over all evaluated samples. 


For serving evaluation, each method is evaluated over $1,000$ watermarked requests under the same workload. Throughput, GPU KV-cache usage, and prefix-cache hit rate are collected
from standard vLLM runtime statistics with log statistics enabled, while
TTFT and TPOT are measured using the standard vLLM benchmark
(\texttt{vllm bench serve}). For asynchronous co-serving systems such as \textsc{SWIFT}, these engine-level metrics may reflect both text and watermark workloads; we therefore treat throughput and E2E TTFT mainly as supplementary serving metrics.


\textbf{Statistical Significance.} To assess runtime stability and repeatability, we repeat the \textsc{SWIFT} experiment over 30 independent runs under the same setting (Table~\ref{tab:latency_repeatability}). \textsc{SWIFT} achieves an average latency of $0.9039 \pm 0.0291$s, with a narrow 95\% confidence interval of $[0.8930\text{s}, 0.9148\text{s}]$. A one-sample $t$-test against the reported latency of $0.905$s gives $p=0.834$, indicating no statistically significant difference from the reported value and confirming stable runtime performance.

\begin{table}[H]
\centering
\small
\caption{Repeatability of \textsc{SWIFT} latency over 30 independent runs. Utility, detectability, and peak memory remain unchanged under the deterministic setting; latency varies due to system runtime.}
\begin{tabular}{lc}
\toprule
\textbf{Metric} & \textbf{Latency (s)} \\
\midrule
Number of runs & 30 \\
Mean $\pm$ Std. & $0.9039 \pm 0.0291$ \\
95\% CI & $[0.8930, 0.9148]$ \\
One-sample $t$-test vs. $0.905$s & $p=0.834$ \\
\bottomrule
\end{tabular}
\label{tab:latency_repeatability}
\end{table}

\begin{figure}[H]
\centering
\subfigure[Text generation]{
\includegraphics[scale=0.25]{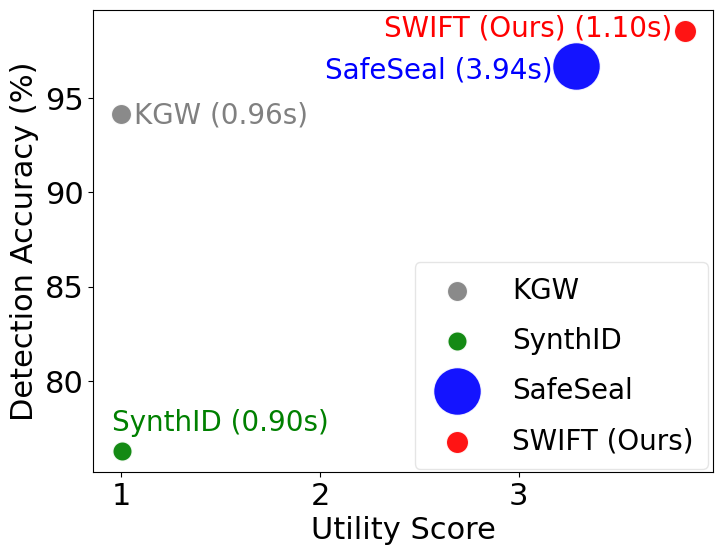}
\label{fig:text_generation_qwen}
} \hspace{1.5cm}
\subfigure[Text summarization]{
\includegraphics[scale=0.25]{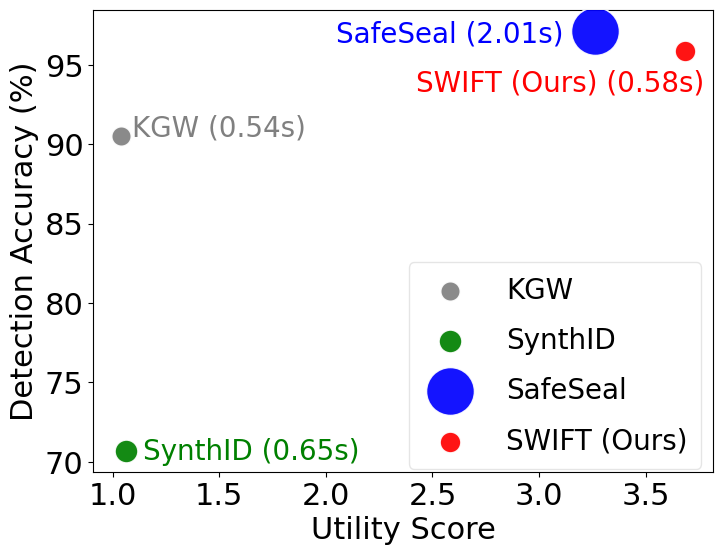}
\label{fig:text_summarization_qwen}}
\caption{Watermark effectiveness on text generation and summarization on Qwen model. Circle
area is proportional to latency, where smaller circles indicate lower latency.}
 \vspace{-7.5pt}
\label{fig:LLM_WM_qwen}
\end{figure}

\textbf{Watermark Effectiveness across LLMs.}
 We further evaluate \textsc{SWIFT} on Qwen2.5-32B \citep{Yang2024Qwen25TR}. As shown in Fig.~\ref{fig:LLM_WM_qwen}, a similar trend is observed for both text generation and summarization. On text generation, \textsc{SWIFT} achieves the highest utility (3.837) and detectability (98.56\%) with competitive latency of 1.102s. On summarization, it again attains the highest utility (3.685) with 95.9\% detectability at low latency of 0.58s, substantially faster than SafeSeal. These results indicate that \textsc{SWIFT}'s effectiveness and efficiency generalize beyond the primary LLM.

\textbf{Performance on Medical Text.} Table~\ref{tab:medmcqa_watermark} provides detailed results on MedMCQA. \textsc{SWIFT} preserves the non-watermarked accuracy of 62.6\% ($\Delta$Acc.=0) while achieving 4.70 utility and 96.70\% detection at 0.401\,s latency. SynthID and SafeSeal also preserve accuracy but achieve lower utility, while KGW slightly reduces accuracy. Although SafeSeal reaches the highest detection rate (97.90\%), it incurs substantially higher latency and lower utility. These results show that \textsc{SWIFT} maintains task correctness while balancing utility, detectability, and serving efficiency on constrained medical text.

\begin{table}[H]
\centering
\small
\caption{Performance on medical MedMCQA dataset. Accuracy is measured against the ground-truth answer, while $\Delta$Acc. denotes the relative accuracy change to non-watermarked baseline.}
\label{tab:medmcqa_watermark}

\begin{tabular}{lccccc}
\toprule
\textbf{Setting}
& \textbf{Accuracy (\%) $\uparrow$}
& \textbf{$\Delta$Acc. (\%) $\uparrow$}
& \textbf{Utility $\uparrow$} 
& \textbf{Detection (\%) $\uparrow$}
& \textbf{Latency (s) $\downarrow$} \\
\midrule

No Watermark
    & 62.6
    & --
    & --
    & --
    & 0.205 \\

KGW
    & 62.3
    & -0.48
    & 2.76
    & 60.80
    & 0.597 \\

SynthID
    & \textbf{62.6}
    & \textbf{0.00}
    & 2.97
    & 66.25
    & 0.416 \\

SafeSeal
    & \textbf{62.6}
    & \textbf{0.00}
    & 3.14
    & \textbf{97.90}
    & 1.732 \\

\textsc{SWIFT} (Ours)
    & \textbf{62.6}
    & \textbf{0.00}
    & \textbf{4.70}
    & 96.70
    & \textbf{0.401} \\

\bottomrule
\end{tabular}
\end{table}

\begin{figure}[H]
\centering
\subfigure[ROC Analysis]{
\includegraphics[scale=0.28]{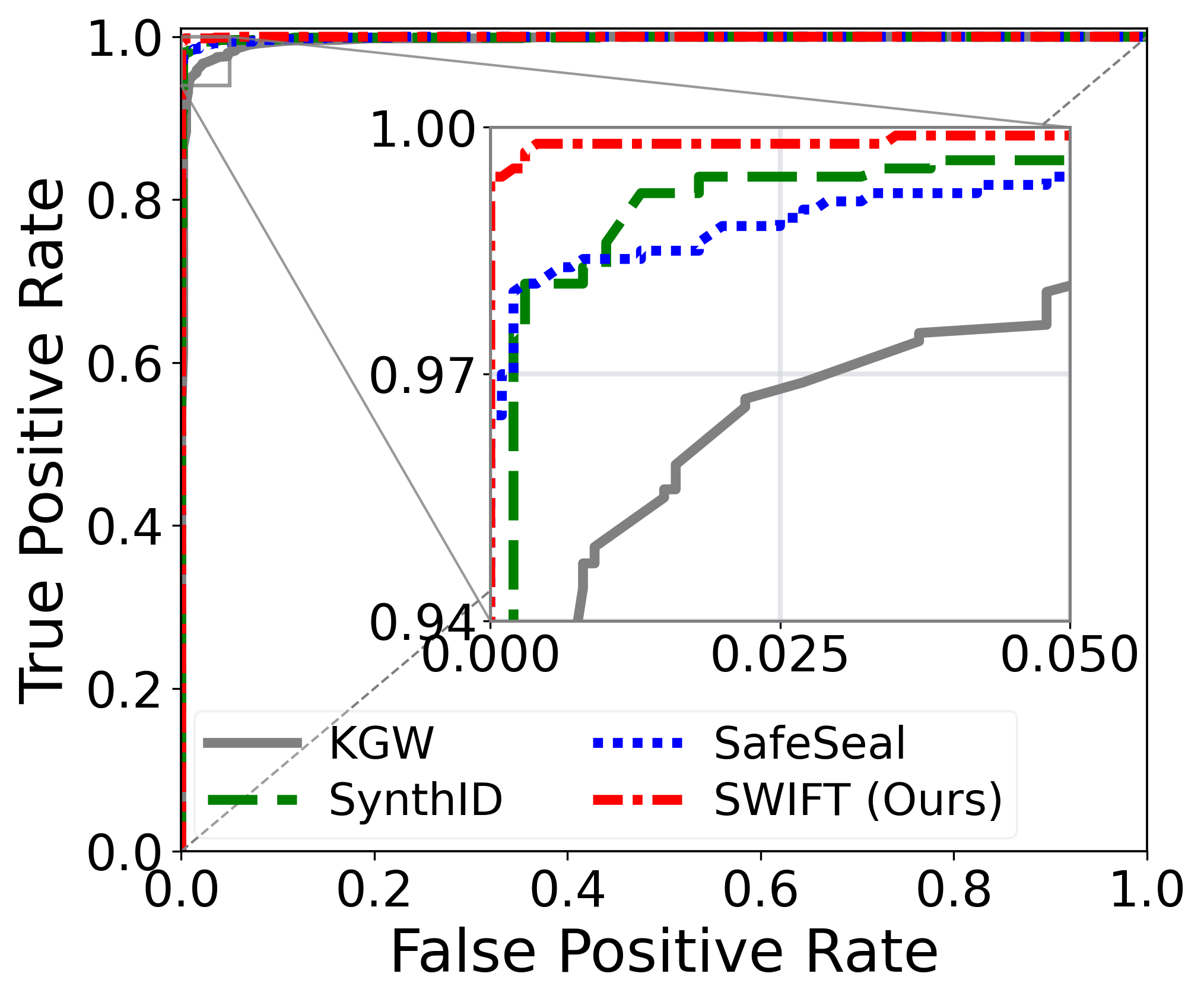}
\label{fig:ROC_curve}
}
\subfigure[Scalability]{
\includegraphics[scale=0.37]{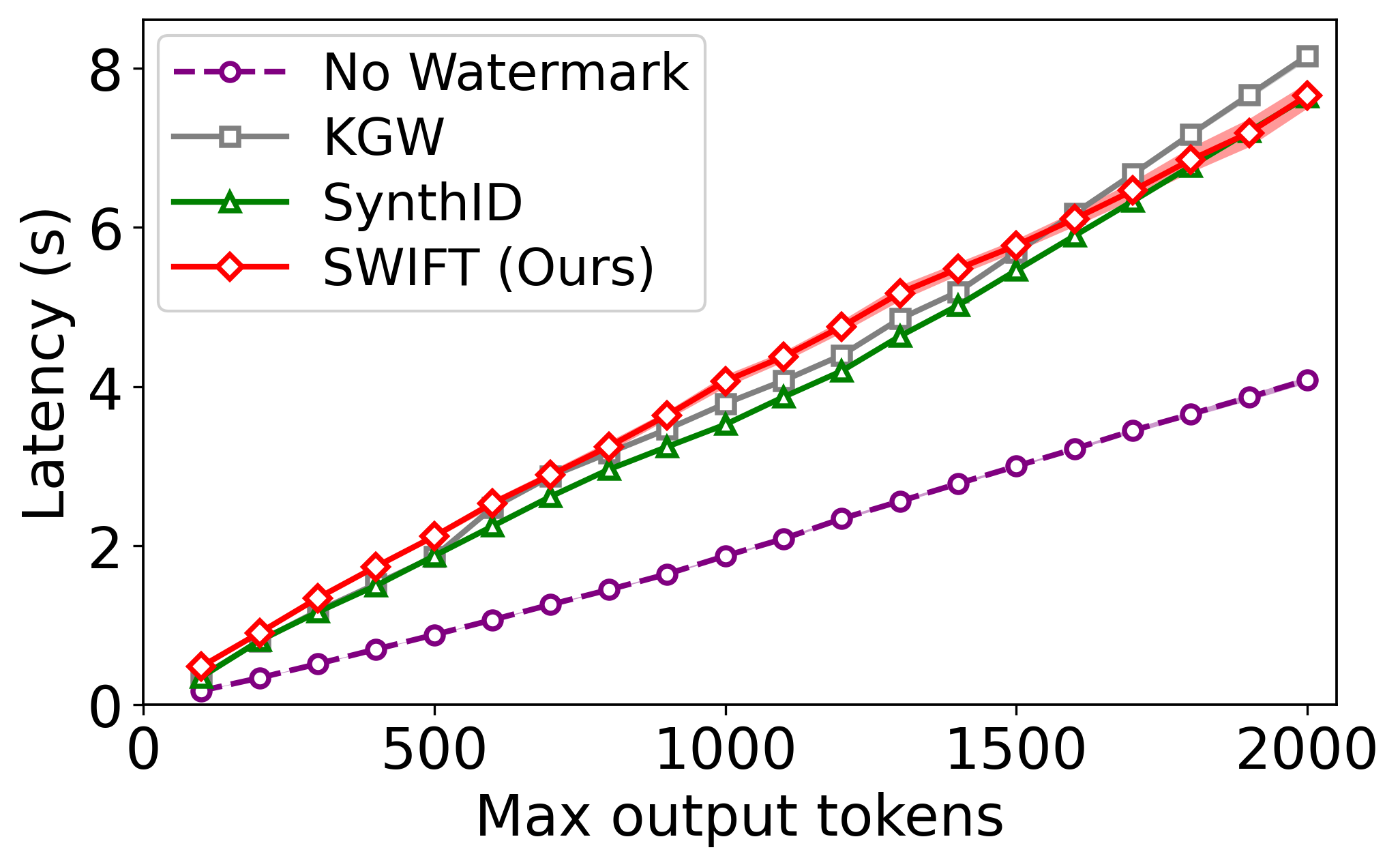}
\label{fig:scalability}
}
\caption{ROC analysis and latency scalability over output length.}
\label{fig:roc_scalability}
\end{figure}

\textbf{ROC Analysis.} 
Fig.~\ref{fig:ROC_curve} further confirms the strong detection performance of \textsc{SWIFT}. It achieves an AUC of 0.9999, compared with 0.9960 for KGW, 0.9987 for SynthID, and 0.9989 for SafeSeal. At a strict 1\% FPR, \textsc{SWIFT} maintains a 99.8\% TPR, increasing to 99.9\% at 5\% FPR. These results indicate strong separation between watermarked and non-watermarked outputs, even at low false-positive operating points. Notably, SynthID achieves a high AUC despite its lower detection accuracy in Section~\ref{sec:main_results}. This difference arises because AUC is threshold-independent and measures the overall separation between the two classes, whereas detection accuracy depends on the selected decision threshold. Overall, all methods perform strongly in the attack-free setting, with \textsc{SWIFT} achieving the highest AUC and maintaining high TPR at low FPR.

\textbf{Scalability over Text Length.} 
Fig.~\ref{fig:scalability} evaluates end-to-end latency as the maximum output length increases from 100 to 2,000 tokens, with least-squares fits showing strong linearity ($R^2 \geq 0.994$). Over this $20\times$ increase in length, \textsc{SWIFT}'s latency increases by only $15.9\times$, indicating less-than-proportional growth over the evaluated range. Its fitted slope is 3.697 ms/token, comparable to SynthID (3.728) and KGW (3.953), demonstrating competitive scaling with output length. SafeSeal is omitted from the plot because its substantially higher baseline latency obscures comparison among the remaining methods. \textsc{SWIFT}'s scaling behavior is supported by asynchronous co-serving, reuse of the fixed watermark instruction prefix, and selective watermarking over only a subset of watermarkable words within each completed segment. In contrast, in-processing methods such as KGW and SynthID apply watermarking decisions at every autoregressive decoding step. Since \textsc{SWIFT}'s cost depends on content-specific factors such as sentence structure and the number of watermarkable positions, we interpret the results as showing favorable scaling comparable to token-level in-processing methods rather than universally superior scaling.

\textbf{Task Characteristics and Watermarking Opportunities.} 
Table~\ref{tab:task_characteristics} summarizes the output characteristics and watermarking opportunities across tasks. We define watermark density as the fraction of output words selected for watermarking. C4 text generation produces substantially longer outputs with more sentences and watermarkable words, resulting in more segment-level watermark requests and greater watermarking workload. In contrast, GSM8K reasoning and Medical QA outputs are shorter and provide more constrained substitution choices, resulting in fewer watermark requests and lower watermark density.

\begin{table}[H]
\centering
\small
\caption{Output characteristics and watermarking opportunities across evaluation tasks. Watermark density denotes the percentage of output words selected for watermarking.}
\label{tab:task_characteristics}
\begin{tabular}{lrrrrrr}
\toprule
\textbf{Dataset}
& \makecell{\textbf{Average}\\\textbf{Tokens}}
& \makecell{\textbf{Average}\\\textbf{Words}}
& \makecell{\textbf{Average}\\\textbf{Sentences}}
& \makecell{\textbf{Watermarked}\\\textbf{Words}}
& \makecell{\textbf{Watermark}\\\textbf{Density (\%)}}
& \makecell{\textbf{Detectability}\\\textbf{(\%)}}\\
\midrule
Text Generation
& 197.74
& 143.89
& 10.37
& 58.29
& 40.51
& 99.65 \\

Summarization
& 61.30
& 46.59
& 3.11
& 16.77
& 35.99
& 98.80 \\

Reasoning
& 48.03
& 38.51
& 3.76
& 8.56
& 22.23
& 90.70 \\

Medical QA
& 83.58
& 59.24
& 3.23
& 17.02
& 28.73
& 96.70 \\
\bottomrule
\end{tabular}
\end{table}

These characteristics help explain the task-dependent latency of \textsc{SWIFT}. Because \textsc{SWIFT} issues watermark requests only for completed sentence segments and selectively modifies watermarkable words, short and constrained outputs generate fewer watermark requests and consume less capacity
in the shared serving batch. In contrast, in-processing methods like KGW and SynthID perform watermarking during autoregressive decoding for every token. This helps explain why \textsc{SWIFT} achieves lower latency than KGW and SynthID on reasoning and medical text (Tables~\ref{tab:gsm8k_watermark} and~\ref{tab:medmcqa_watermark}), with the largest advantage on reasoning, which has the lowest watermark density. The same characteristics also help explain detectability. Lower watermark density provides fewer opportunities to embed the watermark signal, leading to lower detection rates on the more constrained tasks. As shown in Table~\ref{tab:task_characteristics}, detectability generally increases with watermark density, from 90.70\% on reasoning text with 22.23\% density to 99.65\% on C4 with 40.51\% density.

\textbf{Additional Qualitative Examples with Utility Evaluation.}  

\textbf{\textit{On Medical Question-Answering Text.}} Table~\ref{tab:llm_judge_example} presents the LLM-as-a-judge utility evaluation on a MedMCQA example. Although all methods preserve the correct answer, KGW, SynthID, and SafeSeal introduce explanatory, semantic, or factual distortions, receiving utility scores of 3, 2, and 2, respectively. In contrast, \textsc{SWIFT} preserves key medical facts, entities, and relations through meaning-preserving substitutions, achieving the maximum utility score of 5.

\begin{table}[H]
\small
\centering
\caption{Qualitative example of the LLM-as-a-judge utility evaluation on MedMCQA. Green marks \hl{preserved facts}, light blue marks \hlb{meaning-preserving changes}, and red marks \hlred{semantic or factual distortions}. Utility is scored from 0--5 based on semantic and factual preservation.}
\label{tab:llm_judge_example}

\begin{adjustbox}{width=1.0\textwidth}
\begin{tabular}{|p{0.095\linewidth}|p{0.56\linewidth}|p{0.075\linewidth}|p{0.045\linewidth}|p{0.17\linewidth}|}
\hline

\centering \textbf{Setting}
&
\centering \textbf{Content}
&
\centering \makecell{\textbf{Final} \\ \textbf{Answer}}
&
\centering \textbf{Utility}
&
\centering \makecell{\textbf{Judge} \\ \textbf{ Explanation}}
\tabularnewline
\hline

\centering
\multirow{3}{*}{\makecell{\textbf{Question}}}
&
\raggedright
Absorption of which of the following drugs is increased after a fatty meal?

\textbf{A.} Amphotericin B,
\textbf{B.} Griseofulvin,
\textbf{C.} Ampicillin,
\textbf{D.} Aspirin
&
\centering N/A
&
\centering N/A
&
\centering N/A
\tabularnewline
\hline

\centering
\multirow{4}{*}{\makecell{\textbf{Ground} \\ \textbf{Truth}}}
&
\raggedright
\textbf{Answer: B. Griseofulvin.} 
The oral administration of a 0.5\,g dose of griseofulvin produces peak plasma concentrations of 1\,$\mu$g/mL in 4 hours. Its absorption is improved when administered with a fatty meal.
&
\centering \textbf{B}
&
\centering N/A
&
\centering N/A
\tabularnewline
\hline

\centering
\multirow{4}{*}{\makecell{\textbf{No} \\ \textbf{watermark}}}
&
\raggedright
\textbf{Answer: B.}
Griseofulvin is a lipophilic antifungal drug. Its absorption from the gastrointestinal tract is poor and significantly enhanced by concurrent intake of a fatty meal, which increases bile secretion and   improves its solubility.
&
\centering \textbf{B}
&
\centering N/A
&
\centering N/A
\tabularnewline
\hline

\centering
\multirow{5}{*}{\makecell{\textbf{KGW}}}
&
\raggedright
\hl{\textbf{Answer: B.}} \hl{Griseofulvin} is a lipophilic antifungal \hlb{medication}. Its absorption from the gastrointestinal tract is significantly enhanced by concurrent intake of a fatty meal, as it increases bile secretion, which \hlb{aids in its dissolution} and absorption. 
\hlred{Amphotericin B has poor oral absorption regardless of food intake, ampicillin absorption is generally unaffected, and aspirin absorption might be slightly decreased with food due to potential gastric irritation.}
&
\centering \hl{\textbf{B}}
&
\centering 3
&
\raggedright
The medical answer is preserved, but some explanatory details differ or are omitted.
\tabularnewline
\hline

\centering
\multirow{5}{*}{\makecell{\textbf{SynthID}}}
&
\raggedright
\hl{\textbf{Answer: B.}} \hl{Griseofulvin} is a lipophilic antifungal drug \hlred{with poor water solubility}. Its absorption from the gastrointestinal tract is significantly enhanced by concurrent \hlb{ingestion} with a fatty meal, \hlred{as the fat facilitates its dissolution and transport across the intestinal membrane}. 
\hlred{Amphotericin B absorption is actually decreased with food, while ampicillin and aspirin absorption are generally not   affected by dietary fat content.}
&
\centering \hl{\textbf{B}}
&
\centering 2
&
\raggedright
The answer matches, but important explanatory medical content is substantially changed or missing.
\tabularnewline
\hline

\centering
\multirow{4}{*}{\makecell{\textbf{SafeSeal}}}
&
\raggedright
\hl{\textbf{Answer: B.}} \hl{Griseofulvin} is a lipophilic antifungal \hlb{pill}. Its \hlred{emission} from the gastrointestinal tract is \hlred{severe} and \hlb{severely strengthened} by concurrent \hlb{ingest of a fatty plate}, which alters bile \hlred{removal} and   improves its solubility.
&
\centering \hl{\textbf{B}}
&
\centering 2
&
\raggedright
Core meaning is broadly preserved, but several substitutions alter important medical wording and relations.
\tabularnewline
\hline

\centering
\multirow{4}{*}{\makecell{\textbf{\textsc{SWIFT}} \\ \textbf{(Ours)}}}
&
\raggedright
\hl{\textbf{Answer: B.}} \hl{Griseofulvin} is a lipophilic antifungal drug. Its absorption from the gastrointestinal tract is \hlb{inadequate} and significantly \hlb{improved} by \hlb{simultaneous} intake of a fatty meal, which \hlb{raises} bile secretion and  \hlb{facilitates} its solubility.
&
\centering \hl{\textbf{B}}
&
\centering \textbf{5}
&
\raggedright
Semantically equivalent; key facts and entities are preserved.
\tabularnewline
\hline

\end{tabular}
\end{adjustbox}
\end{table}

\textbf{\textit{On Short Reasoning Text.}} 
Table~\ref{table:gsm8k_example} provides a GSM8K example illustrating the effect of watermarking on reasoning quality and the corresponding LLM-as-a-judge utility assessment. KGW and SynthID alter the arithmetic reasoning and produce incorrect final answers, receiving utility scores of 1. SafeSeal preserves the correct answer but introduces semantically inappropriate substitutions such as ``lights'' and ``seats,'' resulting in a utility score of 3. In contrast, \textsc{SWIFT} preserves the reasoning process, quantities, relations, and final answer while applying meaning-preserving lexical substitutions, receiving the maximum utility score of 5. This example further demonstrates \textsc{SWIFT}'s ability to maintain task correctness and semantic fidelity under short reasoning text.

\begin{table}[H]
\small
\centering
\caption{Math reasoning utility example on GSM8K comparing watermarked outputs with the original response. Green marks \hl{correct reasoning and factual consistency}, light blue marks \hlb{semantic-preserving word changes}, and red marks \hlred{reasoning errors or meaning-altering modifications}.}
\begin{adjustbox}{width=1\textwidth}
\begin{tabular}{|p{0.1\linewidth}|p{0.56\linewidth}|p{0.075\linewidth}|p{0.05\linewidth}|p{0.212\linewidth}|}
\hline

\multicolumn{5}{|p{1\linewidth}|}{
\textbf{Question:} 
Hannah has a mental breakdown while studying for finals and starts smashing windows in the high school's parking lot. 
She smashes a quarter of the students' cars' windows and 3/4ths of the teachers' cars' windows. 
If there are 64 students' cars with four windows each and 32 teachers' cars with two windows each, 
how many windows does Hannah smash?
}
\tabularnewline \hline

\centering \textbf{Method} &
\centering \textbf{Reasoning Answer} &
\centering \textbf{Final Answer} &
\centering \textbf{Utility} &
\centering \textbf{Judge Explanation}
\tabularnewline \hline

\centering
\multirow{3}{*}{\makecell{\textbf{No} \\ \textbf{Watermark}}} &
\raggedright
Hannah smashes $64 / 4 * 4 = 16 * 4 = 64$ student car windows. 
Hannah smashes $32 / 4 * 2 * 3 = 8 * 2 * 3 = 48$ teacher car windows. 
Hannah smashes $64 + 48 = 112$ windows.
Answer: 112
&
\centering \hl{\textbf{112}}
&
\centering N/A
&
\centering N/A
\tabularnewline \hline

\centering
\multirow{4}{*}{\makecell{\textbf{KGW}}} &
\raggedright
\hl{Hannah} smashes \hl{$64 / 4 * 4 =$} \hlred{16} students' car windows. 
\hl{Hannah} smashes \hl{$32 / 4 * 2 * 3 =$} \hlred{24} teachers' car windows.
In total, \hl{Hannah} smashes \hlred{$16 + 24 = 40$ windows.} 
\hlred{Answer: 40}
&
\centering \hlred{\textbf{40}}
&
\centering 1
&
\raggedright
The final answer changes, so the conclusion is contradictory.
\tabularnewline \hline

\centering
\multirow{3}{*}{\makecell{\textbf{SynthID}}} &
\raggedright
\hl{Hannah} smashes \hl{$64 / 4 * 4 = 16 * 4 = 64$} student car windows.
\hl{Hannah} smashes \hlred{$32 / 4 * 2 =$ $8 * 2 = 16$} teacher car windows. 
\hl{Hannah} smashes \hlred{$64 + 16 = 80$ windows.} 
\hlred{Answer: 80}
&
\centering \hlred{\textbf{80}}
&
\centering 1
&
\raggedright
The final conclusion changes, creating a contradiction in an important fact.
\tabularnewline \hline

\centering
\multirow{3}{*}{\makecell{\textbf{SafeSeal}}} &
\raggedright
\hl{Hannah} smashes \hl{$64 / 4 * 4 = 16 * 4 = 64$} student car \hlred{lights}. 
\hl{Hannah} \hlb{knocks} \hl{$32 / 4 * 2 * 3 = 8 * 2 * 3 = 48$} teacher car \hlred{seats}. 
\hl{Hannah} smashes \hl{$64 + 48 = 112$ windows.} 
\hl{Answer: 112}
&
\centering \hl{\textbf{112}}
&
\centering 3
&
\raggedright
The quantities and final answer are preserved, but key objects are changed, altering important semantic content.
\tabularnewline \hline

\centering
\multirow{4}{*}{\makecell{\textbf{\textsc{SWIFT}} \\ \textbf{(Ours)}}} &
\raggedright
\hl{Hannah} \hlb{breaks} \hl{$64 / 4 * 4 = 16 * 4 = 64$} \hlb{pupil} car windows. 
\hl{Hannah} \hlb{breaks} \hl{$32 / 4 * 2 * 3 = 8 * 2 * 3 = 48$} \hlb{instructor} car windows.
\hl{Hannah} \hlb{breaks} \hl{$64 + 48 = 112$} windows. 
\hl{Answer: 112}
&
\centering \hl{\textbf{112}}
&
\centering \textbf{5}
&
\raggedright
Meaning, quantities, relations, and key facts are preserved.
\tabularnewline \hline

\end{tabular}
\end{adjustbox}
\label{table:gsm8k_example}
\end{table}


\textbf{Additional Key Serving Evaluation.}

\textbf{\textit{Serving Performance on Text Generation.}}
Table~\ref{tab:serving_efficiency} demonstrates \textsc{SWIFT}'s serving efficiency. \textsc{SWIFT} achieves the highest throughput at 528.7 tokens/s, corresponding to $2.64\times$ KGW, $2.26\times$ SynthID, and $14.1\times$ SafeSeal. It also reduces GPU KV-cache usage to 12.8\%, about 19.5\% lower than KGW and SynthID, while achieving a 96.8\% prefix-cache hit rate compared with nearly 0\% for the baselines. This high hit rate results from \textsc{SWIFT}'s repeated watermark instruction prefix, allowing vLLM to reuse matching KV-cache blocks across watermark requests. In contrast, baseline C4 prompts share few identical full cache blocks (16 tokens by default), yielding little prefix reuse \citep{vllm_cache_block}. Prefix reuse both avoids redundant prefill computation and reduces duplicate KV-cache allocation; additionally, watermark requests have short variable suffixes, limiting their incremental KV-cache footprint. Despite its substantially higher throughput, \textsc{SWIFT} matches the peak memory usage of KGW and SynthID and uses approximately 8.5\% less memory than SafeSeal by sharing a single unified model across text generation and watermark construction.

\begin{table}[H]
\centering
\small
\caption{Serving efficiency comparison across watermarking methods.}
\label{tab:serving_efficiency}

\begin{tabular}{lcccc}
\toprule
\textbf{Methods}
& \makecell{\textbf{Throughput} \\ \textbf{(Tokens/s) $\uparrow$}}
& \makecell{\textbf{GPU KV Cache} \\ \textbf{(\%) $\downarrow$}}
& \makecell{\textbf{Prefix Cache} \\ \textbf{Hit Rate (\%) $\uparrow$}} 
& \makecell{\textbf{Peak Memory} \\ \textbf{(MB) $\downarrow$}}
\\
\midrule

KGW
    & 200.3
    & 15.90
    & 0.00 
    & \textbf{74169}\\

SynthID
    & 234.2
    & 15.90
    & 0.00 
    & \textbf{74169}
    \\

SafeSeal
    & 37.4
    & 15.47
    & 0.00 
    & 81100\\

\textsc{SWIFT} (Ours)
    & \textbf{528.7}
    & \textbf{12.80}
    & \textbf{96.80} 
    & \textbf{74169}\\

\bottomrule
\end{tabular}
\end{table}

\textbf{\textit{Serving Performance on Short Reasoning Text.}} Table~\ref{tab:gsm8k_serving} shows that \textsc{SWIFT} maintains strong serving efficiency on short reasoning workloads. \textsc{SWIFT} achieves the highest throughput at 481.8 tokens/s, approximately $2.56\times$ SynthID and $13.43\times$ SafeSeal, while also obtaining the lowest GPU KV-cache usage (12.7\%) and the highest prefix-cache hit rate (96.68\%). Unlike C4, the baselines also achieve high prefix-cache hit rates (about 87\%) because GSM8K requests share a fixed few-shot reasoning prompt to instruct the model to generate short reasoning text with final answer at the end, creating repeated prefixes that span reusable 16-token cache blocks \citep{vllm_cache_block}. \textsc{SWIFT} further benefits from its repeated watermark instruction prefix, yielding additional KV-state reuse beyond the shared task prompt. These results support that \textsc{SWIFT}'s prefix-caching advantage persists even when the underlying task already provides substantial prefix reuse.

\begin{table}[H]
\centering
\small
\caption{Serving performance on GSM8K short reasoning text. Higher throughput and prefix-cache hit rate are preferred, while lower GPU KV-cache usage is better.}
\label{tab:gsm8k_serving}

\begin{tabular}{lcccc}
\toprule
\textbf{Method}
& \makecell{\textbf{Throughput} \\ \textbf{(Tokens/s) $\uparrow$}}
& \makecell{\textbf{Running} \\ \textbf{Requests} $\uparrow$}
& \makecell{\textbf{GPU KV Cache} \\ \textbf{(\%) $\downarrow$}}
& \makecell{\textbf{Prefix Cache} \\ \textbf{Hit Rate (\%) $\uparrow$}} \\
\midrule

KGW
& 162.35
& 31.09
& 13.46
& 87.59 \\

SynthID
& 187.99
& \textbf{31.56}
& 12.94
& 87.31 \\

SafeSeal
& 35.88
& 31.00
& 16.70
& 87.40 \\

\textsc{SWIFT} (Ours)
& \textbf{481.80}
& 31.10
& \textbf{12.70}
& \textbf{96.68} \\

\bottomrule
\end{tabular}
\end{table}

\textbf{\textit{Additional Serving Evaluation.}} Table~\ref{tab:ttft_tpot} reports Service TTFT, E2E TTFT, and TPOT as supplementary serving measurements. These metrics are primarily defined for token-streaming autoregressive serving and are not fully architecture-independent across watermarking methods. In \textsc{SWIFT}, text-generation and watermark-generation requests are co-served by the same engine; consequently, aggregate benchmark statistics may reflect both request classes rather than isolating user-facing text-generation latency alone. Moreover, \textsc{SWIFT} emits completed watermarked segments, whereas KGW and SynthID operate directly in the token-generation path. We therefore report these values for reference rather than as the primary serving comparison. For SafeSeal, only Service TTFT is reported: its generate-then-post-process design does not expose the token-level timestamps required to obtain comparable E2E TTFT and TPOT from the vLLM serving benchmark.

\begin{table}[H]
\centering
\small
\caption{Serving latency comparison. Service TTFT measures time to first token after request admission, while E2E TTFT includes end-to-end request waiting and serving time. TTFT and TPOT are reported as mean values, where lower is better.}
\label{tab:ttft_tpot}

\setlength{\tabcolsep}{5pt}
\begin{tabular}{lccc}
\toprule
\textbf{Method}
& \textbf{Service TTFT (ms) $\downarrow$}
& \textbf{E2E TTFT (ms) $\downarrow$}
& \textbf{TPOT (ms) $\downarrow$} \\
\midrule

No Watermark
& 128.25
& 140,015.59 
& 46.63 \\
\midrule
KGW
& 495.70
& 441,509.15 
& 172.79  \\

SynthID
& \textbf{412.50}
& \textbf{284,021.78} 
& 91.52  \\

SafeSeal
& 5,342.00
& N/A
& N/A \\

\textsc{SWIFT}
& 483.36
& 359,221.32 
& \textbf{59.65 } \\


\bottomrule
\end{tabular}
\end{table}

\section{Templates} \label{appd:template}
\textbf{Instruction Prefix.} Fig.~\ref{fig:instruction_prompt} provides a template of the shared instruction prefix used in \textsc{SWIFT} for candidate generation. The template includes semantic preservation constraints, entity-aware guidance, general generation rules, and few-shot examples that steer the LLM toward producing context-aware watermark candidates while maintaining consistency with the original text.

\begin{figure*}[!htbp]
\centering
\small

\begin{tcolorbox}[
    enhanced,
    width=1.0\linewidth,
    colback=swiftgray,
    colframe=black!55,
    boxrule=0.5pt,
    arc=2mm,
    left=1.5mm,
    right=1.5mm,
    top=1.5mm,
    bottom=1.5mm
]

\centering
{\bfseries Instruction-Guided Candidate Generation Prompt}

\vspace{0.6em}

\begin{swiftbox}{Shared Instruction Prefix}{swiftblue}
\raggedright
The shared prefix defines the reusable candidate-generation policy for all
sentences in the batch.

\vspace{0.4em}

\begin{tabularx}{\linewidth}{>{\bfseries}p{0.23\linewidth} X}
Target rule &
Select only watermarkable words, which can be naturally replaced, such as nouns, verbs, adjectives, and adverbs. \\

Named entity protection &
Exclude named entities, including names, organizations,
products, titles, dates, quantities, identifiers, and model numbers.

Avoid capitalized words. \\

General Instruction &
Generate 2 - 8 direct synonym candidates that preserve meaning, factuality,
tone, tense, number, degree, and register. 

Candidates can be a single word or a phrase.

Do not generate awkward, duplicate, loose, broader,
narrower, or antonym candidates. \\

Few-shot examples &
\textbf{Example 1:}

\textbf{\textit{Sentence:}} On Tuesday, Dr. Sarah Chen from Stanford University delivered a clear presentation about climate policy, and the audience responded warmly to her practical recommendations.

\textbf{\textit{Output:}} \{"clear":["concise","understandable","lucid"],

"presentation":["talk", "lecture","briefing"],

"responded":["reacted","replied"], 

"warmly":["enthusiastically","favorably"],

"practical":["useful","realistic","workable"],

"recommendations":["suggestions","proposals","advice"]\}

\textbf{Example 2:} \textit{(additional examples omitted for brevity)}

\textbf{Example 3:} \textit{(additional examples omitted for brevity)}
\end{tabularx}
\end{swiftbox}

\vspace{0.45em}

\begin{swiftbox}{Varying Sentence Suffix}{swiftorange}
\raggedright
\textbf{Sentence:}
\texttt{\{completed sentence from asynchronous text generation stream\}}
\end{swiftbox}

\vspace{0.45em}

\begin{swiftbox}{Structured Output}{swiftgreen}
\raggedright
\texttt{\{}\\
\hspace*{1em}\texttt{"handle": ["manage", "oversee", "take care of"],}\\
\hspace*{1em}\texttt{"detailed": ["thorough", "comprehensive"]}\\
\texttt{\}}

\vspace{0.2em}
\textit{Output only a valid JSON object; no markdown, explanation, or extra text.}
\end{swiftbox}

\end{tcolorbox}

\caption{Instruction prompt format for candidate generation in \textsc{SWIFT}. The shared instruction prefix contains the watermarkable word detection, named entity rules, general instructions, and few-shot examples. The varying suffix contains the current sentence, enabling many sentence-level requests to reuse the same shared prefix during continuous batching.}
\label{fig:instruction_prompt}
\end{figure*}

\textbf{LLM-as-a-Judge Utility.} Fig.~\ref{template:llm-judge} shows the utility evaluation prompt. The judge compares each watermarked output with its original response and assigns a score from 0 to 5 based on semantic and factual preservation, including named entities, numbers, relations, actions, and key details. The prompt prioritizes content over stylistic similarity and requires structured JSON output with a score and brief justification. We report the average score across evaluated samples.

\begin{figure*}[t]
\centering
\small

\begin{tcolorbox}[
    enhanced,
    width=1.0\linewidth,
    colback=swiftgreen!10,
    colframe=black!55,
    boxrule=0.5pt,
    arc=2mm,
    left=1.5mm,
    right=1.5mm,
    top=1.5mm,
    bottom=1.5mm
]

\textbf{LLM-as-a-Judge Utility Evaluation Prompt}

\begin{verbatim}
SYSTEM_PROMPT = """
You are an expert evaluator of semantic and factual consistency.

Given an original text (Text A) and a watermarked text (Text B),
evaluate how well Text B maintains the core meaning and factual
content represented in Text A.

Text B does not need to reproduce the exact wording, sentence structure,
or level of detail of Text A. Different but semantically compatible
expressions or formulations should not be penalized.

Consider:
1. Overall semantic meaning and consistency.
2. Named entities, including people, organizations, locations,
   products, titles, and technical terms.
3. Whether important information is contradicted, omitted, or
   materially changed.
4. Numbers, dates, quantities, units, relations, actions, roles,
   and polarity.

Scoring rubric:
0 = Completely unrelated, contradictory, or the core meaning is not preserved.
1 = Same general topic, but most important meaning or facts differ.
2 = Partial semantic consistency, with substantial important information
    changed, contradicted, or missing.
3 = Mostly maintains the core meaning, but some important details,
    entities, or facts differ.
4 = Maintains the core meaning and factual content, with only minor
    semantic or factual differences.
5 = Semantically and factually consistent; key entities, numbers,
    relations, and important information are maintained, even if the
    wording, structure, or presentation differs.

Important:
- If Text B directly contradicts important information in Text A,
  assign a score of at most 1.
- Do not penalize Text B simply for using different wording, sentence
  structure, phrasing, or level of detail.
- Do not require lexical or stylistic similarity between the two texts.
- Focus on substantive semantic and factual differences rather than
  surface-form differences.
- Treat Text A as a reference for semantic and factual consistency,
  not as a template that Text B must closely reproduce.

Return only a JSON object in the following format:
{
  "score": <integer from 0 to 5>,
  "reason": "<brief justification>"
}
"""

USER_PROMPT = """
Text A:
{original_text}

Text B:
{watermarked_text}
"""
\end{verbatim}

\end{tcolorbox}

\caption{Prompt template used for LLM-as-a-judge utility evaluation.}
\label{template:llm-judge}
\end{figure*}

\end{document}